\documentclass[12pt,a4paper]{article}
\pdfoutput=1
\usepackage{jheppub}

\usepackage{bm}
\usepackage{ifsym}
\usepackage{amsthm}
\usepackage{amsfonts}
\usepackage{ccaption}
\usepackage{mathrsfs}
\usepackage{booktabs}
\usepackage{enumerate}

\makeatletter
\@addtoreset{equation}{section}

\makeatletter
\renewcommand\section{\@startsection {section}{1}{\z@}%
                                   {-3.5ex \@plus -1ex \@minus -.2ex}
                                   {2.3ex \@plus.2ex}%
                                   {\normalfont\large\bfseries}}
\renewcommand\subsection{\@startsection{subsection}{2}{\z@}%
                                     {-3.25ex\@plus -1ex \@minus -.2ex}%
                                     {1.5ex \@plus .2ex}%
                                     {\normalfont\bfseries}}

  \captionnamefont{\bfseries}
  \captiontitlefont{\small\sffamily}
  \captiondelim{: }
  \hangcaption

\def\sec#1{\S\ref{#1}}
\def\fig#1{Fig.\,\ref{#1}}
\def\req#1{(\ref{#1})}

\def\RR{{\mathbf R}}

\definecolor{rust}{rgb}{0.8,0.2,0.2}
\definecolor{green}{rgb}{0.1,0.8,0.2}
\definecolor{hue55}{rgb}{0.0,0.5,0.67}
\definecolor{hue75}{rgb}{0.65,0.15,0.6}

\def\AdS#1{AdS$_{#1}$}
\def\SAdS#1{Schwarzschild-AdS$_{#1}$}

\def\bdy{{\cal B}}

\def\SAdS{Schwarzschild-AdS}

\def\rh{r_h}   
\def\hh{\mu}
\def\rhotp{\rho_0}  
\def\rhot{{\tilde \rho}}  
\def\tA{t_{{\cal A}}}   
\def\phA{\varphi_{{\cal A}}}  

\def\cCA{{\cal A}^c}

\def\CA{{\cal A}}
\def\CB{{\cal B}}

\def\CH{{\cal H}}
\def\RR{{\mathbf R}}
\def\ph{\varphi}

\def\ddtipp{{q^\vee}}  
\def\ddtipf{{q^\wedge}}  
\def\cwedge{\blacklozenge_{{\cal A}}} 
\def\CW#1{\blacklozenge_{#1}}  
\def\bCW#1{\partial \CW{#1}} 
\def\bCWA{\partial \cwedge} 
\def\domd{\Diamond_{\cal A}}  

\def\lcrit{\ell_\ast}
\def\phAcrit{\varphi^\ast_{{\cal A}}}
\def\rhocrit{\rho_\ast}

\def\bdy{\partial{\cal M}}

\def\curf#1{\gamma_{#1}^+}
\def\curp#1{\gamma_{#1}^-}

\def\entsurf{\partial {\cal A}}  

\def\bcwedgef{\partial_+(\blacklozenge_{{\cal A}})}
\def\bcwedgep{\partial_-(\blacklozenge_{{\cal A}})}

\def\csfz#1{{ \Psi}_{#1}}   
\def\extr#1{{\mathfrak E}_{#1}}   
\def\csf#1{{\Xi}_{#1}}  

\title{Global properties of causal wedges in asymptotically AdS spacetimes}

\author{Veronika E. Hubeny$^a$}
\author{  Mukund Rangamani$^a$}
\author{\& Erik Tonni$^b$}

\affiliation[a]{ Centre for Particle Theory \& Department of Mathematical Sciences,\\
Science Laboratories, South Road, Durham DH1 3LE, UK.}
\affiliation[b]{SISSA and INFN, via Bonomea 265,  34136, Trieste, Italy}
\emailAdd{veronika.hubeny@durham.ac.uk}
\emailAdd{mukund.rangamani@durham.ac.uk}
\emailAdd{erik.tonni@sissa.it}

\abstract{We examine general features of causal wedges in asymptotically AdS spacetimes and show that in a wide variety of cases they have non-trivial topology. We also prove some general results regarding minimal area surfaces on the causal wedge boundary and thereby derive constraints on the causal holographic information. We go on to demonstrate that certain properties of the causal wedge impact significantly on features of extremal surfaces which are relevant for computation of holographic entanglement entropy.} 


\begin{document}
\begin{flushright} \small{DCPT-13/25} \end{flushright}

\maketitle

\flushbottom

\section{Introduction \& summary}
\label{s:intro}

One of the fundamental properties of a Lorentzian spacetime is its causal structure. In the context of the holographic AdS/CFT correspondence, the importance of causality was noted very early on \cite{Horowitz:1999gf}  and it was realized that bulk AdS causality should be at the very least compatible with boundary causality. In fact in asymptotically  AdS spacetimes,  causal propagation through the bulk cannot be faster than propagation along the boundary as a consequence of the gravitational time-delay effect \cite{Woolgar:1994ar, Gao:2000ga}. Nevertheless, we do not understand the holographic dictionary well enough to pin-point a particular feature of the boundary field theory data that we can directly associate with the bulk causal structure. It is however clear that for any boundary field theory state which is described by a semi-classical bulk geometry, one can use certain observables of the field theory to probe aspects of the bulk causal structure, as has been noted in various contexts in the past, see e.g.\ \cite{Kabat:1999yq,Balasubramanian:1999zv,Louko:2000tp,Gregory:2000an, Marolf:2004fy,Hubeny:2005qu, Hubeny:2006yu}.

Of interest to us is the set of bulk spacetime points which is naturally associated with a particular spatial region of the boundary field theory. This question has been tackled in a number of different ways in the recent past \cite{Bousso:2012sj, Czech:2012bh, Hubeny:2012wa, Bousso:2012mh}. The various constructions described in these works can be divided into two classes: those that use regions bounded in the bulk by extremal surfaces inspired by the holographic entanglement entropy  \cite{Ryu:2006ef} and its covariant generalization \cite{Hubeny:2007xt}, and those that put causal relations at the center stage. From these analyses and earlier works \cite{Hubeny:2012ry} the following picture emerges: a given spatial region ${\cal A}$, whilst certainly being cognizant of the causal wedge associated with it in the bulk, is however able to access information of part of the geometry beyond the causal wedge.

Nevertheless, as noted in  \cite{Bousso:2012sj, Hubeny:2012wa}, one might say that the most natural bulk spacetime region associated with a spatial region $\CA$ on the boundary is the causal wedge $\cwedge$ and associated quantities, since these are constructed solely using causal relations. Furthermore, the causally inspired constructions,  as we shall see, serve to bound other observables, and in a certain sense the causal wedge is the bare minimum that the boundary field theory region should reproduce.
Inspired by this observation, we undertake an examination of causal wedges in asymptotically AdS spacetimes (see \cite{Ribeiro:2005wm, Ribeiro:2007hv} for some previous observations).  Despite being a simple exercise, it reveals rather interesting surprises; in fact we will demonstrate that the causal structure constrains other observables such as the entanglement entropy in a non-trivial fashion.

\subsection{Causal constructions: a review}
\label{s:causalrev}

To set the stage for our discussion let us quickly recall some basic concepts relevant for the causal constructions. 
Consider a ($d$-dimensional) boundary spacetime foliated by a set of ($d-1$ dimensional) Cauchy slices $\Sigma_t$, labeled by boundary time $t$. We will consider spacelike  ($d-1$ dimensional) regions $\CA_t , \CB_t , \ldots \in \Sigma_t$ (generically we drop the subscript $t$ for notational simplicity). The complement of a region ${\cal A}$ will be denoted as $\cCA$.

\begin{figure}
\begin{center}
\includegraphics[width=5in]{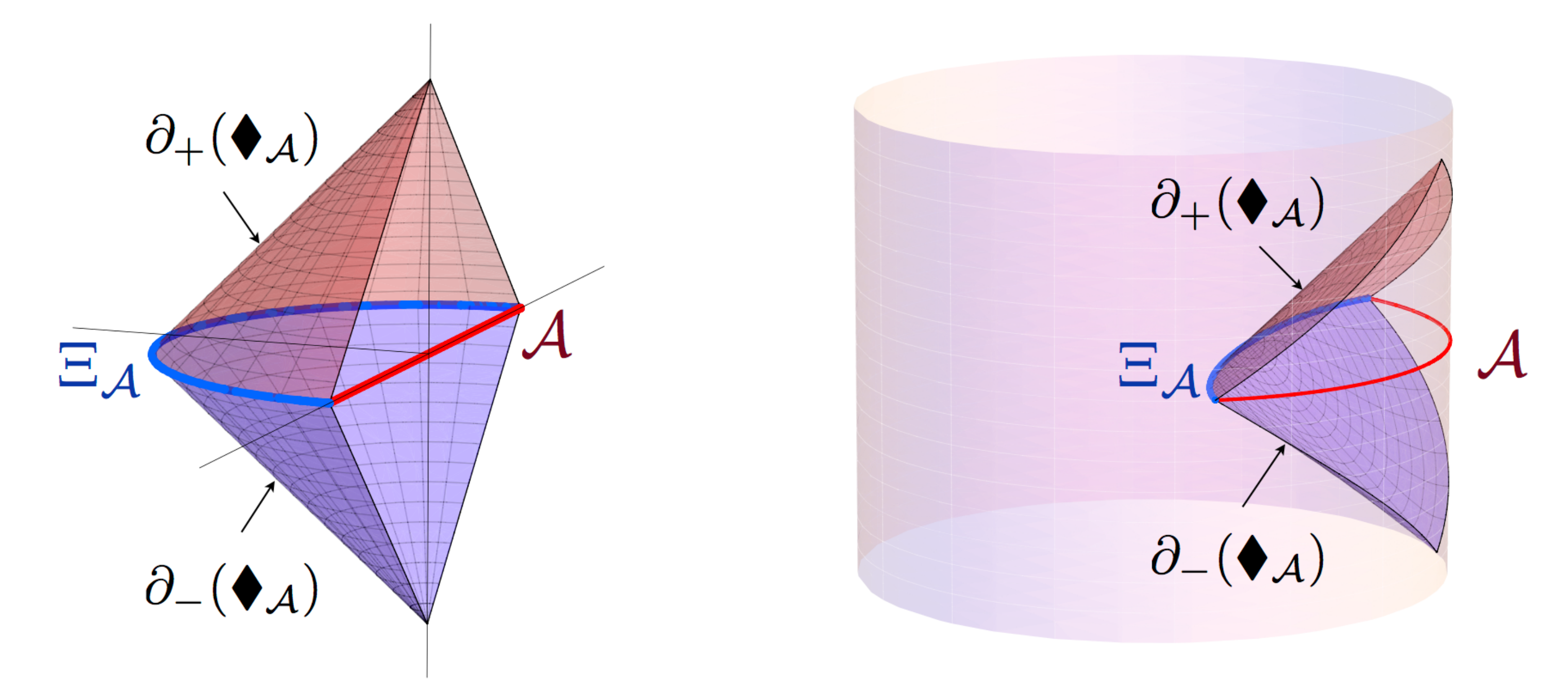} 
\caption{
A sketch of the causal wedge $\cwedge$ and associated quantities in planar AdS (left) and global AdS (right) in 3 dimensions taken from \cite{Hubeny:2013hz}: in each panel, the region $\CA$ is represented by the red curve on right, and the corresponding surface $\Xi_\CA$ by blue curve on left; the causal wedge $\cwedge$ lies between the AdS boundary and the null surfaces $\bcwedgef$ (red surface) and $\bcwedgep$ (blue surface).}
\label{f:CWsketch}
\end{center}
\end{figure}

 In a nutshell, the {\it causal wedge} $\cwedge$, associated to a boundary region $\CA$, is the set of bulk spacetime points which lie in both the future and the past of the boundary domain of dependence\footnote{
The domain of dependence $\domd$ corresponds to the boundary spacetime region whose physics is fully determined by initial conditions at $\CA$.  More formally, any fully extended timelike curve on the boundary which passes through $\domd$ must necessarily intersect $\CA$.
} $\domd$ for the region $\CA$,
\begin{equation}
\cwedge \equiv J^-[\domd] \cap  J^+[\domd] \ .
\label{}
\end{equation}	
In other words, the causal wedge consists of the set of spacetime events through which there exists a causal curve which starts and ends in $\domd$.
The boundary of $\cwedge$ restricted to the bulk, denoted $\bCWA$, consists of two\footnote{
There may in fact be multiple boundaries; but we postpone a discussion of this subtlety till \sec{s:CWtopol}.
} null surfaces  $\partial_\pm (\cwedge)$ which are generated by null geodesics; the outgoing null geodesics ending on the future boundary of $\domd$ generate $\bcwedgef$ and the ingoing ones from past boundary of $\domd$ generate $\bcwedgep$.
These bulk co-dimension one null surfaces
$\bcwedgef$ and $\bcwedgep$ intersect along a bulk co-dimension two spacelike surface $\csf{\cal A}$, which for reasons explained in   \cite{Hubeny:2012wa} we dub the {\it causal information surface}.  In other words,
\begin{equation}
\bCWA 
	 = \bcwedgef \cup \bcwedgep 
\qquad {\rm and} \qquad
\csf{\cal A}
	 = \bcwedgef \cap \bcwedgep \ .
\label{defXi}
\end{equation}
By construction, the surface $\csf{\cal A}$ is anchored on the entangling surface $\partial {\cal A}$ of the selected region ${\cal A}$, i.e.\ $\partial(\csf{\cal A})  = \partial {\cal A} $.
For orientation, these constructs are illustrated in \fig{f:CWsketch}, for planar AdS (left) and global AdS (right).
We emphasize that our construction is fully general and covariant, requiring only a spacetime geometry that allows us to define causal curves.  For convenience we focus on a causal spacetime with a smooth metric.

Given the bulk co-dimension two surface $\csf{\cal A}$, one can associate a scalar quantity with the region ${\cal A}$. 
In analogy with the definition of the holographic entanglement entropy \cite{Ryu:2006bv,Ryu:2006ef,Hubeny:2007xt}, in  \cite{Hubeny:2012wa} we defined the {\it causal holographic information} of $\CA$, abbreviated $\chi_\CA$,  as quarter of the proper area of $\Xi_\CA$ in Planck units,
\begin{equation}
\chi_{\cal A} \equiv \frac{\text{Area}(\csf{\cal A})}{4\,G_N} \ .
\label{}
\end{equation}	
Although this number is infinite since $\Xi_\CA$ reaches to the AdS boundary, as with entanglement entropy, there is information both in the divergence structure as well as in the regulated quantity.  

The simplicity of these bulk constructs suggests that they should have correspondingly natural field theory dual.  Moreover, the importance of causal structure in bulk gravity indicates that the dual field theory constructs should likewise correspond to  fundamental quantities. 
In order to obtain hints of what these field theoretic quantities are, we set out to explore the bulk properties of the causal wedge and associated constructs.  In \cite{Hubeny:2012wa} we have considered certain static geometries in order to study its properties in equilibrium, and suggested that $\chi_\CA$ captures the basic amount of information about the bulk geometry contained in the reduced density matrix $\rho_\CA$ for the region $\CA$.  
While $\chi_\CA$ provides a (rather weak upper) bound on the entanglement entropy $S_\CA$, for certain special cases -- which happen to be the ones where we can actually calculate the entanglement entropy from first principles -- the causal information surface $\Xi_\CA$ in fact coincides with the extremal surface $\extr{\CA}$, so this bound is saturated: $\chi_\CA = S_\CA$.
To glean more intuition about the dynamical nature of our constructs, in \cite{Hubeny:2013hz} we focused specific time-dependent backgrounds corresponding to rapid thermalization. We observed that in general, if we consider a region $\CA$ at a certain boundary time $\tA$, the construction of $\Xi_\CA$ is quasi-teleological (i.e.\ teleological only on light-crossing timescale) in time-dependent backgrounds, since the domain of dependence $\domd$ contains times later than $\tA$.\footnote{
Curiously, while $\chi$ behaves correspondingly quasi-teleologically in general time-dependent backgrounds, in the case of collapsing thin null shell examined in \cite{Hubeny:2013hz}, the temporal evolution of $\chi$ remains entirely causal.
In other words, while the causal information surface $\Xi$ gets deformed by the shell quasi-teleologically, its area $\chi$ remains unaffected till after the region $\CA$ itself encounters the shell.} Finally, the recent analysis of \cite{Freivogel:2013zta} attempts to provide some interesting speculations about the field theoretic interpretation of $\chi_{\cal A}$. 

In the present work, we continue the exploration of properties of causal wedge and related constructs, now in complete generality.  This presents a somewhat complementary approach to that of \cite{Hubeny:2013hz}: instead of focusing on specific class of spacetimes where the explicit computation of $\chi_\CA$ is tractable, we maximally relax the assumptions about the bulk spacetime, and consider  global properties that these constructs, especially $\cwedge$ and $\Xi_\CA$, must satisfy in general.  Nevertheless, to exemplify our statements and familiarize the reader with the subtleties, we will present explicit constructions of $\Xi_\CA$ in a particular spacetime in \sec{s:CWtopol}.  

\subsection{Preview of results}
\label{s:preview}

We will start in \sec{s:CWtopol} by examining the topological structure of the causal wedge $\cwedge$ for a simply connected boundary region $\CA$. 
A similar problem was considered in \cite{Ribeiro:2007hv}, with the conclusion that for simply connected regions $\CA$, the causal wedge must likewise be simply connected. This follows from the statement of topological censorship \cite{Galloway:1999bp}, which asserts that every causal curve which begins and ends on the boundary of asymptotically AdS spacetime ${\mathscr I} = \partial \text{AdS}$ is contractible to the boundary. Viewing the domain of dependence $\domd \subset {\mathscr I}$ as a particular causal sub-domain, the result follows. 

Here we will point out that the situation is actually more subtle.  
Despite the simplicity of $\CA$, the associated causal wedge may be topologically arbitrarily complicated: for example, it can have (arbitrarily) many holes, i.e., non-trivial homology. In the examples we encounter, we will show that in asymptotically \AdS{d+1} bulk spacetime, it is possible for $\partial\cwedge$ to possess non-contractible co-dimension two spheres ${\bf S}^{d-1}$. Thus $\Xi_\CA$ may be composed of multiple disconnected components. Moreover, the change in topology can be engineered by varying 
parameters associated with ${\cal A}$ relative to those characterizing the bulk geometry, implying interesting `phase transitions' for $\chi$. 

This can be observed already in perhaps the simplest non-trivial example distinct from pure AdS, namely a neutral static black hole in global AdS$_{d+1}$ for $d>2$, provided $\CA$ is large enough, as explained in \sec{s:SAdS} (cf.\ \fig{f:CWplot} for an illustrative plot).
Despite the simplicity of the setup, this result may come as a surprise to many readers, and in fact (to our knowledge) has not been pointed out previously in the literature.  There may be several reasons for this.  Initial studies of the causal wedge (such as explicitly carried out in \cite{Hubeny:2013hz}) often focus on 3-dimensional bulk geometry, where this effect is absent.  For example, the BTZ black hole causal wedge is simply connected for any region size, as we explicitly illustrate in \sec{s:BTZ}.  Moreover, most higher-dimensional studies typically focus on planar black holes, so as to consider duals of states of CFT on Minkowski background.  However, the effect we describe is  absent for a planar black hole, since there is no region `on the other side' of the black hole which could be causally accessible from the boundary. We illustrate explicitly that when the CFT lives on the Einstein static universe, $\csf{\cal A}$ is disconnected whenever the region ${\cal A}$ covers a sufficient portion of the full system (ranging from more than half of the full system in case of tiny black holes to almost the entire system in case of huge black holes).

Considering large regions $\CA$ (i.e.\ comparable to the size of the full system on compact space) may perhaps seem too artificial to worry about.  To dispel such objections, we proceed in \sec{s:ConfSol} to argue that in fact the causal wedge for arbitrarily small region can also have holes, for a suitably chosen bulk geometry.  A simple example is provided by the global completion of the  conformal soliton \cite{Friess:2006kw}, which is in fact just a boosted version of a static black hole \cite{Horowitz:1999gf}.
One can also consider  genuinely dynamical situations involving multiple black holes. Though the explicit metric is not known analytically, such configurations provide simple existence arguments for causal wedges with multiple holes (and correspondingly multiple components of a disconnected $\csf{\cal A}$).  

Based on the above examples, one might easily wonder if the non-trivial structure of a causal wedge is somehow inherited from the bulk geometry being causally non-trivial.  It is in fact easy to argue that this is {\it not} the case, as we discuss in \sec{s:genAdS}.  A causally trivial spacetime, such as a compact star, can likewise admit causal wedges with holes.  This further solidifies the robustness of this feature.

Having seen  in \sec{s:CWtopol} that even in simple bulk spacetimes causal wedges may have surprising properties, the reader might be led to suspect that this will render the CFT dual of the causal wedge and derived quantities far too complicated.  However, we take the viewpoint that since the causal wedge $\cwedge$ is the simplest and most natural bulk construct associated with a boundary region $\CA$, it ought to have a natural CFT dual nevertheless.
In \sec{s:chigen}  we collect the global properties that any causal wedge must satisfy.  These include natural inclusion properties, as well as simple additivity properties of the causal holographic information $\chi$, which are previewed in \sec{s:properties} and justified in \sec{s:proofs}.

Having established these global properties for the constructs ($\cwedge$, $\Xi_\CA$, and $\chi_\CA$) derived from our causal wedge, in \sec{s:extrsurf} we turn to a brief discussion of implications specifically for the extremal surface $\extr{\CA}$ and the entanglement entropy $S_\CA$ associated to a given boundary region $\CA$.  Most intriguingly, the property established in \sec{s:chigen} that any extremal surface $\extr{\CA}$ must lie outside\footnote{
By outside we mean here that no part of the extremal surface $\extr{\CA}$ can lie within the causal wedge $\cwedge$.} 
of the causal wedge $\CA$, together with the observation of \sec{s:CWtopol} that  the causal wedge has a hole (i.e.\ $\Xi$ consists of two disconnected components) in the \SAdS\ background for sufficiently large region $\CA$, implies that for such cases, there does not exist a connected extremal surface anchored on $\partial \CA$ homologous to $\CA$.
In particular, the connected extremal surface $\extr{\cCA}$ corresponding to the complement $\cCA$ of the region $\CA$ does not satisfy the homology requirement since it is separated from $\CA$ by the black hole, while a surface going around the black hole so as to be homologous to $\CA$ would necessarily have to enter inside the causal wedge.
This means that the extremal surface whose area gives the entanglement entropy $\extr{\CA}$ must likewise consist of two disconnected components: one given by $\extr{\cCA}$ and the other wrapping the black hole horizon.  That in turn implies that for such cases (i.e.\ sufficiently large $\CA$), the difference in entanglement entropies $S_{\cal A}- S_{\CA^c}$ captures precisely the thermal entropy $S_{BH}$, and therefore saturates the Araki-Lieb inequality \cite{Araki:1970ba}.\footnote{
This observation is further discussed in \cite{Hubeny:2013ee} where it is explicitly demonstrated that connected minimal surfaces fail to capture the entanglement entropy for large enough regions ${\cal A}$. The resulting saturation of the Araki-Lieb inequality is referred to there as the entanglement plateau phenomenon.} This is perhaps the most interesting result of our explorations, justifying our intuition that the properties of the causal wedges serve to non-trivially constrain physically understood observables such as entanglement entropy.

Given the potentially profound implications of the existence of causal wedges with holes, as well as the relative ease with which one can construct examples of such an occurrence, one might start to worry that perhaps for any bulk geometry different from pure AdS (and satisfying the null energy condition), we could construct a sufficiently large boundary region $\CA$ whose causal wedge $\cwedge$ has a hole.  We address this possibility in \sec{s:discuss} and suggest that this does not happen unless the deformation is sufficiently strong; in spherically symmetric situations, we conjecture that 
the presence of holes in the causal wedge is associated with the presence of null circular orbits in the spacetime.

\section{Topological structure of causal wedge}
\label{s:CWtopol}

Let us first examine the topological structure of the causal wedge $\cwedge$ for a given bulk spacetime and a  simply connected boundary region $\CA$. 
Recall that the causal wedge can be thought of as consisting of causal curves which begin and end in $\domd$.  Although these curves are all continuously deformable into each other, we show that the causal wedge can be topologically more complicated.  For example, $\Xi_\CA$ need not be homologous to $\CA$, or even when it is, $\bCWA$ need not be homotopic to $\domd$.  We focus on $\Xi_\CA$, specifically whether or not it is connected and the nature of transitions between the number of its components.  

To motivate the possibility of $\Xi_\CA$ having multiple components, let us observe that a causal wedge cannot penetrate a black hole (i.e., $\Xi_\CA$ cannot reach beyond an event horizon). Heuristically this follows directly from the definition of a black hole: no causal curve from inside can reach the AdS boundary, much less $\domd \subset {\mathscr I}$ (see also \sec{s:chigen}).  Suppose however that the black hole is very small (compared to AdS scale).  Far away from the black hole, its gravitational effects are negligible; so there is a spatial region surrounding the black hole from which causal curves which would have reached well within $\domd$ in pure AdS still reach $\domd$ in the actual spacetime.  In other words, although the black hole itself cannot lie inside the causal wedge, a spatial region fully surrounding it is contained in $\cwedge$.
This reasoning suggests that for small black holes and sufficiently large regions $\CA$, the causal information surface $\Xi_\CA$ has two components: one similar to that in pure AdS which is anchored on $\partial \CA$, and one which shields the black hole.  We will now check this expectation explicitly for \SAdS$_{d+1}$ for $d\ge3$ (focusing on $d=4$ which is algebraically simplest) and then comment on other geometries.

\subsection{Global \SAdS$_{d+1}$}
\label{s:SAdS}

To illustrate the point  that even for simply-connected regions $\CA$ the causal information surface $\Xi_\CA$ may be composed of multiple disjoint components, let us consider the \SAdS$_{d+1}$ black hole.  We will w.l.o.g.\ set the AdS scale to unity, and characterize the black hole by its horizon size in AdS units, $\rh \in (0,\infty)$.  This gives a 1-parameter family of static, spherically symmetric and physically well-behaved spacetimes with metric
\begin{equation}
ds^2 = -f(r)\, dt^2 + \frac{dr^2}{f(r)} + r^2 \, \left(d\ph^2 + \sin^2\ph\,  d\Omega_{d-2}^2\right) \,, \qquad f(r) =  r^2+1 - \frac{\rh^{d-2}  \,  (\rh^2 + 1)}{r^{d-2}} \ .
\label{}
\end{equation}	
Large black holes ($\rh > 1$) are dual to the thermal density matrix of the field theory on the Einstein Static Universe  $\text{ESU}_{d} ={\bf S}^{d-1} \times {\mathbb R}$ (e.g.\ as in the case of ${\cal N}=4$ SYM in $d=4$), while small black holes are still physically relevant in the microcanonical ensemble.\footnotemark
\footnotetext{
The translation between field theory and geometry is the following: the black holes with horizon size $\rh$ have a Hawking temperature $T_{BH} = \frac{d\,\rh^2+(d-2)\,\ell_\text{AdS}^2}{4\pi\,\rh\,\ell_\text{AdS}^2}$. These solutions have a minimum value of $T_{BH}$ attained at $\rh = \sqrt{\frac{d-2}{d}} \,\ell_\text{AdS}$. They however minimize the free energy only for $T_{BH} \,\ell_\text{AdS}\geq \frac{d-1}{2\pi}$ or equivalently $\rh \geq \ell_\text{AdS}$.} 
For convenience we will restrict attention to boundary regions $\CA$ which preserve $SO(d-1)$ spherical symmetry.  For purposes of finding the causal wedge $\cwedge$, we can then reduce this problem to effectively 3-dimensional one\footnote{
Note that any curve which is causal in the full $(d+1)$-dimensional space is necessarily causal in the reduced $(2+1)$-dimensional subspace, and conversely any causal curve in the 3-dimensional spacetime trivially lifts to a causal curve in the full $(d+1)$-dimensional spacetime.
} by reducing the ${\bf S}^{d-1}$ to one non-trivial angle $\ph \in [0,\pi]$.  The region $\CA$ is then characterized by its radius $\phA$.
As $\phA \to \pi$, the region covers most of the boundary space.\footnote{
To keep $\domd$ finite, we will however consider $\phA<\pi$ in this section.  (As explained in \sec{s:chigen},
when $\phA = \pi$ the boundary region ${\cal A} = {\bf S}^{d-1}$ is a complete Cauchy slice of the Einstein Static Universe. Then $\cwedge$ is simply the region exterior of the black hole and $\csf{\cal A}$ is the bifurcation surface of the horizon.) \label{fn:horizon}}  On the other hand, the planar black hole case is recovered in the limit $\phA \to 0$ and $\rh \to \infty$.

It is convenient to use coordinates $(t,\rho,\ph)$ where $\rho\in [0,\pi/2)$ is related to the standard radial coordinate $r$ by $r=\tan \rho$.  The relevant 3-dimensional piece of the bulk metric is then
\begin{equation}
ds^2 = \frac{1}{\cos^2 \rho} \left(- g(\rho) \, dt^2 + \frac{d\rho^2 }{g(\rho)} + \sin^2 \rho \, d\ph^2 \right) \ ,
\label{metgofrho}
\end{equation}	
where 
\begin{equation}
g(\rho) = 1 - \hh \, \frac{\cos^d \rho}{\sin^{d-2} \rho} \ , \qquad
\hh \equiv \rh^{d-2} \,  (\rh^2 + 1) = \frac{\sin^{d-2} \rho_h}{\cos^d\rho_h} \ .
\label{}
\end{equation}	
Null geodesics in this subspace are characterized  by the reduced angular momentum $\ell \in (-1,1)$, a discrete parameter $\eta=\pm 1$ labeling outgoing ($\eta = 1$) versus ingoing ($\eta = -1$) geodesics, as well as the initial position.  We can write the differential equations in terms of the affine parameter $\lambda$ where $\dot{} \equiv \frac{d}{d\lambda}$ as follows:
\begin{equation}
{\dot t} = \frac{\cos^2 \rho}{g(\rho)}
\ , \qquad
{\dot \ph} = \ell \, \frac{\cos^2 \rho}{\sin^2 \rho}
\ , \qquad
{\dot \rho} = \eta \, \cos^2 \rho \, \sqrt{1- \ell^2 \,  \frac{g(\rho)}{\sin^2 \rho}} \ .
\label{SAdSgeodap}
\end{equation}	
We could in principle obtain analytic expressions for  $(t(\lambda),\rho(\lambda),\ph(\lambda))$, which are given in terms of elliptic functions. This however does not add much insight and it is easier to see the structure graphically, so we simply integrate the geodesic equations numerically.
In actual implementation it is convenient to solve \req{SAdSgeodap} for $t(\rho)$ and $\phi(\rho)$ directly, though we have to keep track of $\eta$ changing sign at a turning point where ${\dot \rho} = 0$.  

Only geodesics with sufficiently large angular momentum have a turning point; these are ones for which the equation ${\dot \rho} = 0$ in \req{SAdSgeodap} has a real solution $\rho_0 \in (0 , \frac{\pi}{2})$.  It is easy to check that this only occurs for 
$\ell^2 \in \left(\ell_0^2 \ , \  1 \right) $, where
\begin{equation}
\ell_0 = 
\left[ 1+ (d-2) \, d^{-\frac{d}{d-2}} \, \left( \frac{2}{\mu} \right)^{\! \frac{2}{d-2}} \right]^{-1/2}
\qquad \xrightarrow{d=4} \
\sqrt{\frac{4\hh}{1+4\hh}} = \frac{2 \, \rh \, \sqrt{\rh^2+1}}{2\, \rh^2 + 1} \ .
\label{ellOofrh}
\end{equation}	
The corresponding value of $\rho_0$ at this minimal $\ell_0$ corresponds to the circular null orbit radius and is given by
\begin{equation}
\rho_0(\ell_0) = 
\tan^{-1} \left( \frac{d \, \mu}{2} \right)^{\! \frac{1}{d-2}}
\qquad \xrightarrow{d=4} \
\tan^{-1} \sqrt{2 \, \mu} \ .
\label{}
\end{equation}	
For general $\ell>\ell_0$, the radial position of the turning point is given by  the largest root of the polynomial $\dot \rho =0$ which is of order $d$ (or $\frac{d}{2}$ for even $d$); e.g.
\begin{equation}
\rhotp =   \tan^{-1}  \sqrt{\frac{\ell^2}{2 \, (1-\ell^2) } \, \left[ 1+ \sqrt{1-4 \, \hh \, \left( \frac{1-\ell^2}{\ell^2} \right)} \right] } \qquad {\rm for} \ \ d=4 \ .
\label{rhotp}
\end{equation}	
(On the other hand, geodesics with $\ell<\ell_0$ have no turning point: instead they terminate at the curvature singularity at $r=0$.)

Now that we have $\rhotp$, we can find the corresponding $\ph$ and $t$ coordinates for the turning point.
We typically specify the particular $\ell$-geodesic by giving the functions $t_\ell(\rho)$ and $\ph_\ell(\rho)$ and  when necessary explicitly indicating whether we are before or after the turning point. In particular, we have the following expressions for the trajectory of the future-directed null geodesics from a boundary point $\ddtipp$ defined below (hence initially $\eta  = -1$)  parameterized by $\ell$:
\begin{align}
& 
t_\ell(\rho) =  t_i
+\int_{\rho_i}^{\rho_f}  \frac{h_\ell(\rhot)}{g(\rhot)} \, d\rhot   \,, 
\qquad 
\ph_\ell(\rho) = \ph_i+ \ell  \int_{\rho_i}^{\rho_f}\; \frac{h_\ell(\rhot)}{\sin^2\rhot} \, d\rhot  \,, \qquad 
h_\ell(\rho) \equiv \frac{1}{\sqrt{1-\ell^2\, \frac{g(\rho)}{\sin^2\rho}}} \ ,
\nonumber \\ 
& \text{ingoing segment } (\eta  = -1): \quad t_i = t_\vee, \quad \ph_i = 0,\quad \rho_i = \rho , \quad \rho_f = \frac{\pi}{2} \ , \nonumber \\
& \text{outgoing segment } (\eta  = 1): \quad t_i  = t_\text{ingoing}(\rho_0) \equiv t_0, \quad \ph_i =\ph_\text{ingoing}(\rho_0) \equiv \ph_0,\quad \rho_i = \rho_0 ,\quad \rho_f = \rho \ .
\label{trhopieces}
\end{align}	
Similar expressions can be written down for the geodesics that end up on $\ddtipf$.

Let us now turn to the strategy for finding $\Xi_\CA$.
Since the geometry is static, we can w.l.o.g.\ place $\CA$ at time $t=0$.  The domain of dependence $\domd$
for round regions of radius $\phA$ 
 is then determined by two boundary points 
$\domd = J^+[\ddtipp] \cap J^-[\ddtipf]$
with $(t,\rho,\ph)$ coordinates
$\ddtipp = (-\phA, \frac{\pi}{2}, 0)$ and $\ddtipf = (\phA, \frac{\pi}{2},  0)$.\footnote{
We adapt the notation introduced in \cite{Hubeny:2012wa}; for simple regions the causal wedge is generated by null geodesics emanating from two points $\ddtipf$ in the future and $\ddtipp$ in the past.}
Recall that for general spacetimes the boundary of the full causal wedge $\cwedge$ is generated by bulk null geodesics which start at $\ddtipp$ (for $\bcwedgep$) or end on $\ddtipf$ (for $\bcwedgef$).  
Their intersection gives $\Xi_\CA$.  Since the \SAdS\ spacetime is static, the two sets of geodesics are merely time-reversed versions of each other, and their intersection necessarily lies at $t=0$.
Furthermore, by spherical symmetry, each congruence respects $\ph$-reversal symmetry.  This means that it suffices to find just one congruence to determine the rest.  
For convenience of discussion let us label the four sets of geodesics by letters $P$ ($F$) for past (future) congruence and $R$ ($L$) for right (left) part of each congruence, i.e.\ ingoing from the boundary towards positive (negative) $\ph$.  So $PR$ and $PL$ geodesics generate $\bcwedgep$ and $FR$ and $FL$ geodesics generate $\bcwedgef$, with $\ell>0$ along the $PR$ and $FL$ congruence.
Hence we only need to  find the solution to \req{SAdSgeodap} given by $t_\ell(\rho)$ and $\ph_\ell(\rho)$ for the $PR$ congruence, say; the symmetries
\begin{equation}
PR_\ell(t,\rho,\ph) 
 = PL_\ell(t,\rho,-\ph) 
  = FR_\ell(-t,\rho,\ph) 
   = FL_\ell(-t,\rho,-\ph) 
\label{geodreflections}
\end{equation}	
 then immediately give the $PL$, $FR$, and $FL$ congruences.  

Let us now consider the intersections of these congruences.\footnote{
For the \SAdS$_{d+1}$ spacetime, it can be checked 
that geodesics {\it within} each congruence do not intersect each other.
}
For each $\ell$, $PR_\ell$ geodesic intersects $FR_\ell$ geodesic at $t=0$
 and similarly for the $PL_\ell$ and $FL_\ell$ geodesics.  
Let us denote the curve generated by these intersections (i.e.\ parameterized by $\ell$) on the $t=0$ surface  by ${\cal X}_{t=0}$, and let us denote its coordinates by 
$(\rho_{t=0}(\ell),\ph_{t=0}(\ell))$ and $(\rho_{t=0}(\ell),-\ph_{t=0}(\ell))$ for the $R$ and $L$ halves of the congruences respectively.  
For each $\ell$, we can find these by first solving $t_\ell(\rho) = 0$ for $\rho$ and then substituting this into $\ph_\ell(\rho)$ to determine $\ph$.
Consider now the function $\ph_{t=0}(\ell)$.  For the radial geodesics, we necessarily have $\ph_{t=0}(0)=0$, whereas for the boundary geodesics, it is easy to see that $\ph_{t=0}(1)=\phA < \pi$.  
If $\ph_{t=0}(\ell)$ increases monotonically with $\ell$, or more generally if $\ph_{t=0}(\ell) <\pi$ for all $\ell \in (0,1)$, then ${\cal X}_{t=0}$ gives the full curve $\Xi_\CA$.  

However, there exists another set of intersections between the congruences which becomes relevant if $\ph_{t=0}(\ell) >\pi$.  In particular, the $FR_\ell$ geodesic intersects the $FL_\ell$ geodesic at $\ph = \pi$, and similarly for $PR$ and $PL$ geodesics.  When $\ph_{t=0}(\ell) >\pi$, the $PR$ geodesic from $\ddtipp$ (along which $\ph$ starts from 0 and increases monotonically with the affine parameter) would have intersected the $PL$ geodesic (at $\ph = \pi$) before intersecting the $FR$ geodesic (at $t=0$ and $\ph = \ph_{t=0}(\ell) >\pi$).  The moment two null geodesics from $\ddtipp$ intersect,\footnote{
In  slight abuse of language but following previous terminology, we will refer to these intersections as caustics and denote them by ${\cal C}^\pm$ for the $\partial _\pm(\cwedge)$ congruences.
} they henceforth become timelike-separated from $\ddtipp$, and therefore enter inside $\cwedge$,  no longer remaining on $\bcwedgep$.  The subsequent intersection on  
${\cal X}_{t=0}$ is therefore not on $\bcwedgep$ and correspondingly is not relevant for $\Xi_\CA$.  Said differently, $\Xi_\CA$ closes off at $\ph = \pi$.  

To summarize, the condition for $\Xi_\CA$ to have two disconnected components is 
\begin{equation}
\max_{\ell \in (0,1)} \ph_{t=0}(\ell) >\pi \ .
\label{diconnectedXicond}
\end{equation}	
If \req{diconnectedXicond} holds, then there are two\footnote{
A-priori, there could have been an even number larger than 2, but explicit checks indicate that this doesn't happen; essentially there isn't enough structure in the geodesic equations for \SAdS\ to allow multiple extrema.
Said differently, there are two competing effects which influence how much a geodesic `orbits' in a given time span: 
the light bending gets stronger nearer to the black hole, but so does the time-delay.  
To maximize the former while minimizing the latter, we need to tune  $\ell$ to attain the optimal penetration depth (approximately given by the null circular orbit radius); our assertion follows since $ \ph_{t=0}(\ell)$ increases for smaller $\ell$ and decreases for larger $\ell$.
}
solutions of $ \ph_{t=0}(\ell) =\pi$; let us label them by $0 <\ell_1 < \ell_2 <1$.
In such a case, $\Xi_\CA$  has one component given by   ${\cal X}_{t=0}$ for $\ell \in (0,\ell_1)$ and another given by   ${\cal X}_{t=0}$ for $\ell \in (\ell_2,1)$.  
The latter is connected to the AdS boundary and is anchored at $\partial \CA$ for $\ell =1$.  The former is disconnected from the boundary and wraps the black hole.   
\begin{figure}
\begin{center}
\includegraphics[width=5in]{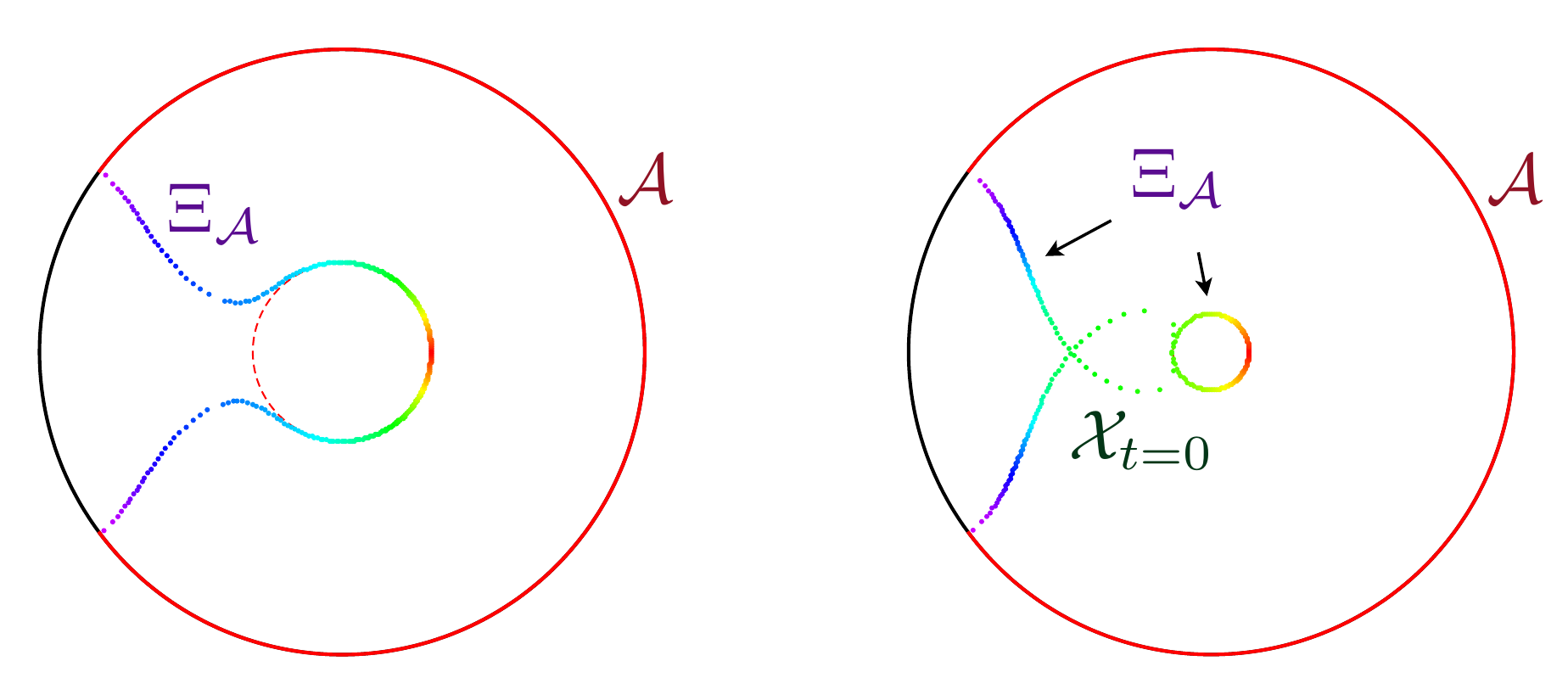} 
\caption{
A plot of the intersection points of the future and past congruence, ${\cal X}_{t=0}$, plotted on the Poincar\'e disk.  In each panel, the outer circle represents the AdS boundary (with the region $\CA$ highlighted in red; $\phA=2.5$ in both panels).  The black hole size is $\rh = 0.5$ (left) and $\rh = 0.2$ (right), denoted by red dashed curve (but obscured in the latter case).
${\cal X}_{t=0}$ is composed of the individual intersection points, color-coded by $\ell$ (from red at $\ell =0$ to purple at $\ell = 1$).  For large enough black hole (left), $ \ph_{t=0}(\ell) <\pi$ for all $\ell$, and therefore ${\cal X}_{t=0} = \Xi_\CA$.  For small black hole (right) ${\cal X}_{t=0}$ self-intersects and therefore $\Xi_\CA$ has two components as indicated.
}
\label{f:Xi_PD}
\end{center}
\end{figure}
This situation is illustrated in \fig{f:Xi_PD}, where we plot ${\cal X}_{t=0}$ on the Poincar\'e disk for connected (left) and disconnected (right) case.  

\begin{figure}
\begin{center}
\includegraphics[width=3in]{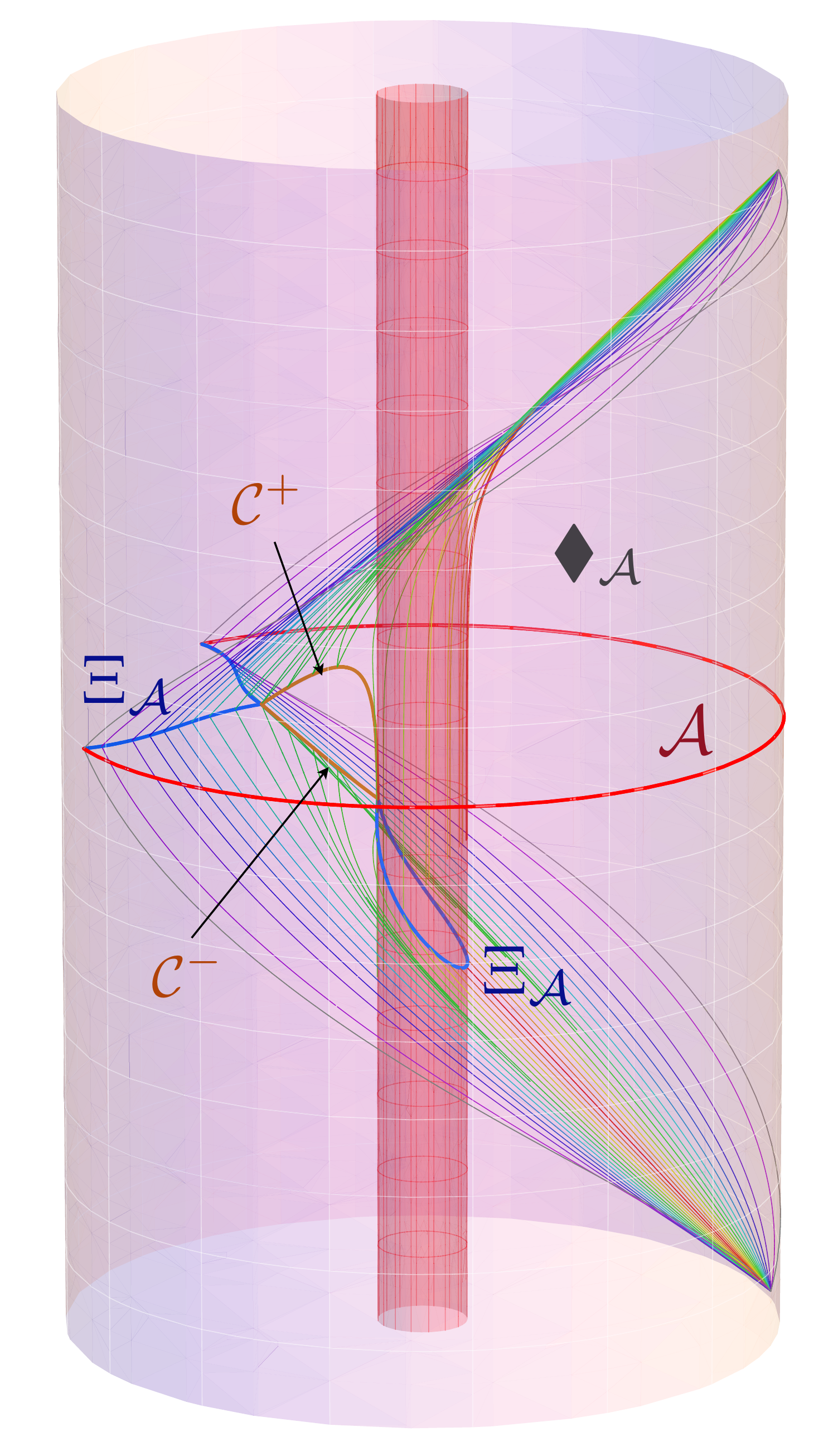} 
\caption{
Causal wedge for the case $\rh=0.2$ and $\phA=2.5$, as in right panel of \fig{f:Xi_PD}.  Same color-coding (by $\ell$) is applied to the null geodesic generators of $\partial_\pm \cwedge$.  In addition to the AdS boundary and horizon, the plot exhibits the region $\CA$  (indicated by the thick red curve), the two components of $\Xi_\CA$  (indicated by the thick blue curves), and the curves of caustics ${\cal C}^\pm$ (indicated by thick brown curves) which connect up the two components of $\Xi_\CA$.
The causal wedge $\cwedge$ bounded by the null generators clearly exhibits a hole.
}
\label{f:CWplot}
\end{center}
\end{figure}

To understand better what happens in the disconnected case, it is instructive to consider the full causal wedge.
This is illustrated in \fig{f:CWplot} where we plot the causal wedge for the same set of parameters as in the right panel of \fig{f:Xi_PD}, but now on a 3-d spacetime diagram in ingoing Eddington coordinates.\footnote{
Following previous convention \cite{Hubeny:2013hz}, we plot $\rho$ radially and choose the vertical coordinate such that ingoing radial null geodesics lie at 45$^\circ$.
This fixes the vertical coordinate to be given by $v-\rho+\frac{\pi}{2}$, where
\begin{equation} \nonumber
v = t+ \frac{1}{2 \rh^2+1} \left( 
	\sqrt{\rh^2+1} \,  \left[ 
	\tan^{-1} \frac{\tan \rho}{\sqrt{\rh^2+1}} - \frac{\pi}{2} \right]
	- \rh \, \tanh^{-1} \frac{\rh}{\tan \rho}
\right) \ .
\end{equation}	
For this reason, the plot is asymmetric under vertical flip.
}
There are several features of note:  as expected, the causal wedge clearly has a hole, causing $\Xi_\CA$ to have two disconnected components, one anchored on $\partial \CA$ and one wrapping the black hole.  This was already necessitated by the observation that the causal wedge cannot penetrate the black hole, while approximating the pure AdS causal wedge far away from the black hole.  However, unlike the pure AdS case, the boundary of the causal wedge has caustics where the $L$ and $R$ geodesics from the same congruence intersect each other (before intersecting those from the other congruence).  The two caustic curves (${\cal C}^+$ on $\bcwedgef$ and ${\cal C}^-$ on $\bcwedgep$) lie at $\ph=\pi$ and connect the two components of $\Xi_\CA$, where the latter cusps.

As an aside, we remark that the presence of caustics in the causal wedge implies that generically the causal information surface $\Xi_\CA$ need not be smooth; for less symmetric spacetimes this can happen even when $\Xi_\CA$ is connected.
Although that might seem like a bizarre feature, it is worth remembering that it is actually no worse than the  analogous property of an event horizon, which  likewise  is not smooth generically.\footnote{
Event horizons are locally Lipschitz, but not necessarily more regular than that \cite{Helfer:2011vv}.
There even exist (rather exotic) examples \cite{Chrusciel:1996tw} where the event horizon is `nowhere' differentiable, in the sense of non-existence of any open neighbourhood where the horizon is differentiable, though they are differentiable `almost everywhere' in a measure-theoretic sense.  More generally, \cite{Beem:1997uv} showed that differentiability fails precisely where new generators join the horizons.}
In particular, although a horizon generator has to remain on the horizon, new generators can enter the horizon at caustics.

Let us now return to considering the effect of varying the parameters, namely the transition between connected and disconnected $\Xi_\CA$ exemplified in \fig{f:Xi_PD}.
As we decrease the black hole size $\rh$ (for a fixed $\phA$), the curve $ \ph_{t=0}(\ell) $ reaches higher and higher, eventually exceeding $\pi$ (and growing without bound as $\rh \to 0$).  This is because the null geodesics whose angular momentum is close to $\ell_0$ orbit around the black hole many ($\sim \rh^{-1}$) times before reaching $t=0$.
Conversely, for fixed $\rh$, the causal information surface $\Xi_\CA$ is connected if $\phA$ is sufficiently small (and is guaranteed to be so if $\phA<\pi/2$ for any $\rh$) and disconnected if $\phA$ is large enough (and is guaranteed to be so when $\phA \to \pi$  for any $\rh$).  
\begin{figure}
\begin{center}
\includegraphics[width=3in]{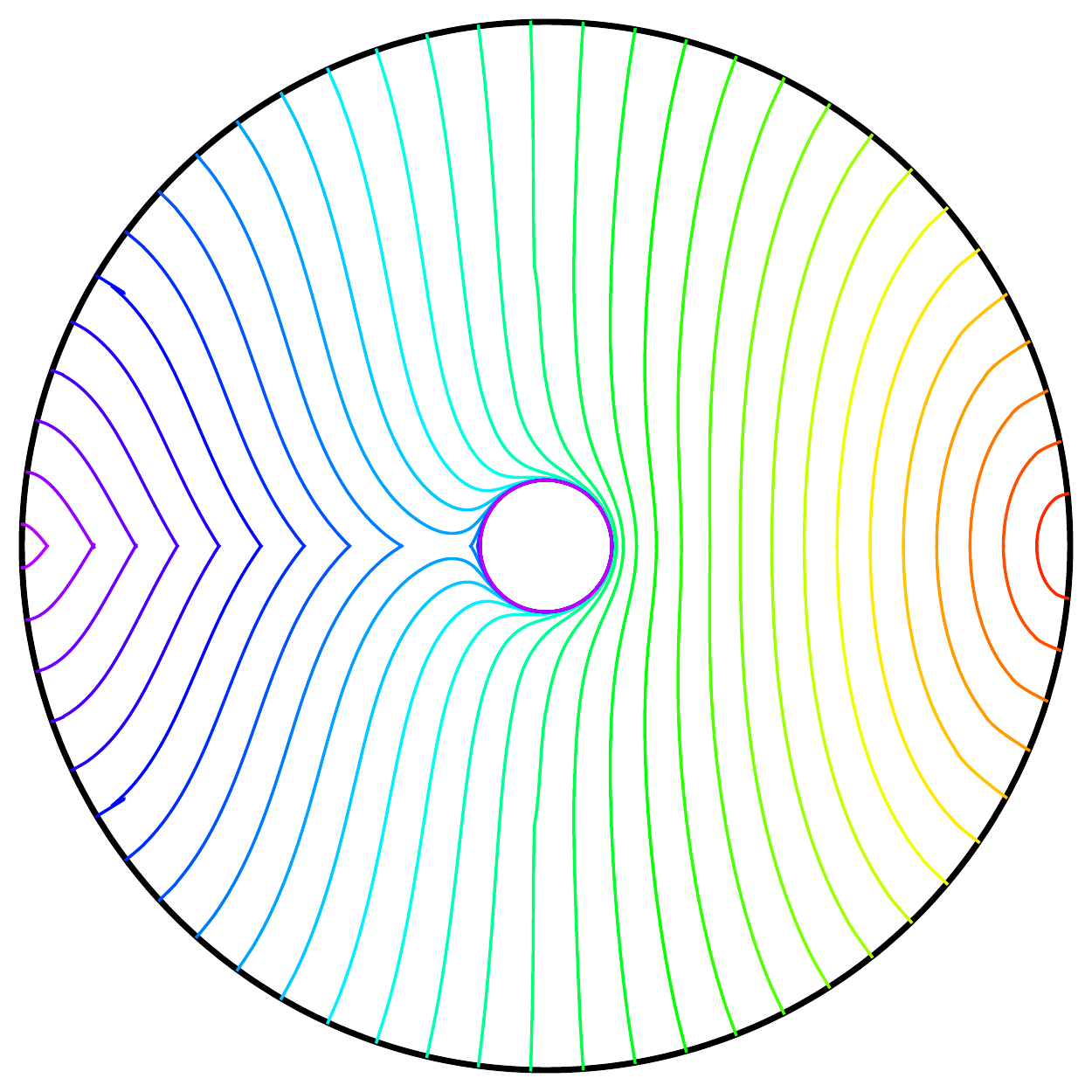} 
\caption{
The projection of the causal information surfaces $\Xi_\CA$ for various $\phA$ onto the Poincar\'e disk of \SAdS$_5$ with fixed black hole size $\rh=0.2$, color-coded by $\phA$ which varies from $0$ (red) to $\pi$ (purple) in increments of $0.1$.  (For example, the blue curve with $\phA=2.5$ corresponds to the projection of $\Xi_\CA$ in \fig{f:CWplot}.)   We can clearly see that $\Xi_\CA$ pinches off; for larger $\phA$, the disconnected component of $\Xi_\CA$ is located very near the horizon.
}
\label{f:PDXifoliation}
\end{center}
\end{figure}
This is illustrated in \fig{f:PDXifoliation}, which plots the causal information surfaces $\Xi_\CA$ for the full range of $\phA$ at a fixed black hole size ($\rh=0.2$, as in \fig{f:CWplot}), projected onto the Poincar\'e disk.   This presents a somewhat complementary information to that in \fig{f:Xi_PD}: whereas the latter varied $\rh$ at fixed $\phA$, \fig{f:PDXifoliation} varies $\phA$ at fixed $\rh$. 

Since $\phA$ is connected for small $\phA$ and disconnected for sufficiently large $\phA$, we're guaranteed by continuity that for any $\rh$, there is a critical region size $\phAcrit$ for which $\Xi$ just pinches off.  
Let us denote the angular momentum along the corresponding null geodesic (i.e.\ the one passing through this pinch-off point) by $\lcrit$.
The four geodesics, 
$PR_{\lcrit}$, $PL_{\lcrit}$, $FR_{\lcrit}$, and  $FL_{\lcrit}$, 
all intersect at a single point $t=0, \ph=\pi$, and $\rho = \rhocrit$.
While this is true whenever $\ph_{t=0}(\ell) = \pi$, here we have an extra condition that $\Xi_\CA$ self-intersects at a tangent, i.e.\ that $\max_{\ell \in (0,1)} \ph_{t=0}(\ell) = \pi$.  This condition is satisfied only for a specific relation between $\rh$ and $\phA$.  
Hence to determine $\lcrit$ and $\rhocrit$, and correspondingly the critical curve in $(\rh,\phA)$ plane, we need to be able to find $\max_{\ell \in (0,1)} \ph_{t=0}(\ell)$ efficiently.  

While doing this numerically is rather time-consuming, we can simplify matters using the following observation.  Geodesics with angular momentum $\ell_0$, when close to $\rho \approx \rho_0$, move very slowly in the $\rho$ direction.  Hence to compensate, i.e.\ to remain null, they have to move faster in the $\ph$ direction to cross the same temporal distance than nearby-$\ell$ geodesics (which move at finite speed in the $\rho$ direction everywhere).  Assuming that this effect dominates over what happens to the geodesics in passage between the boundary and vicinity of $\rho_0(\ell_0)$,
the $\ell_0$ geodesic will reach $\ph=\pi$ `earlier' than the nearby-$\ell$ geodesics.  This in turn means that if we cut them off at $t=0$ to find $\Xi_\CA$, the pinch-off which happens at $\ph = \pi$ will occur near the $\ell_0$ geodesic.
This implies that $\lcrit \approx \ell_0$.
Assuming that in fact $\lcrit = \ell_0$, it is very simple to find the critical curve in $(\rh,\phA)$ plane given by a function $\phA^\ast(\rh)$, since for each fixed $\rh$ (which determines $\ell_0$ using \req{ellOofrh}), we merely need to find $\rhocrit$ by solving $t_{\ell_0}(\rhocrit)=0$, and then integrating the $\ell_0$-geodesic from $(t=0, \rho = \rhocrit, \ph=\pi)$ back out to the boundary to find $\phA$.

\begin{figure}
\begin{center}
\includegraphics[width=5in]{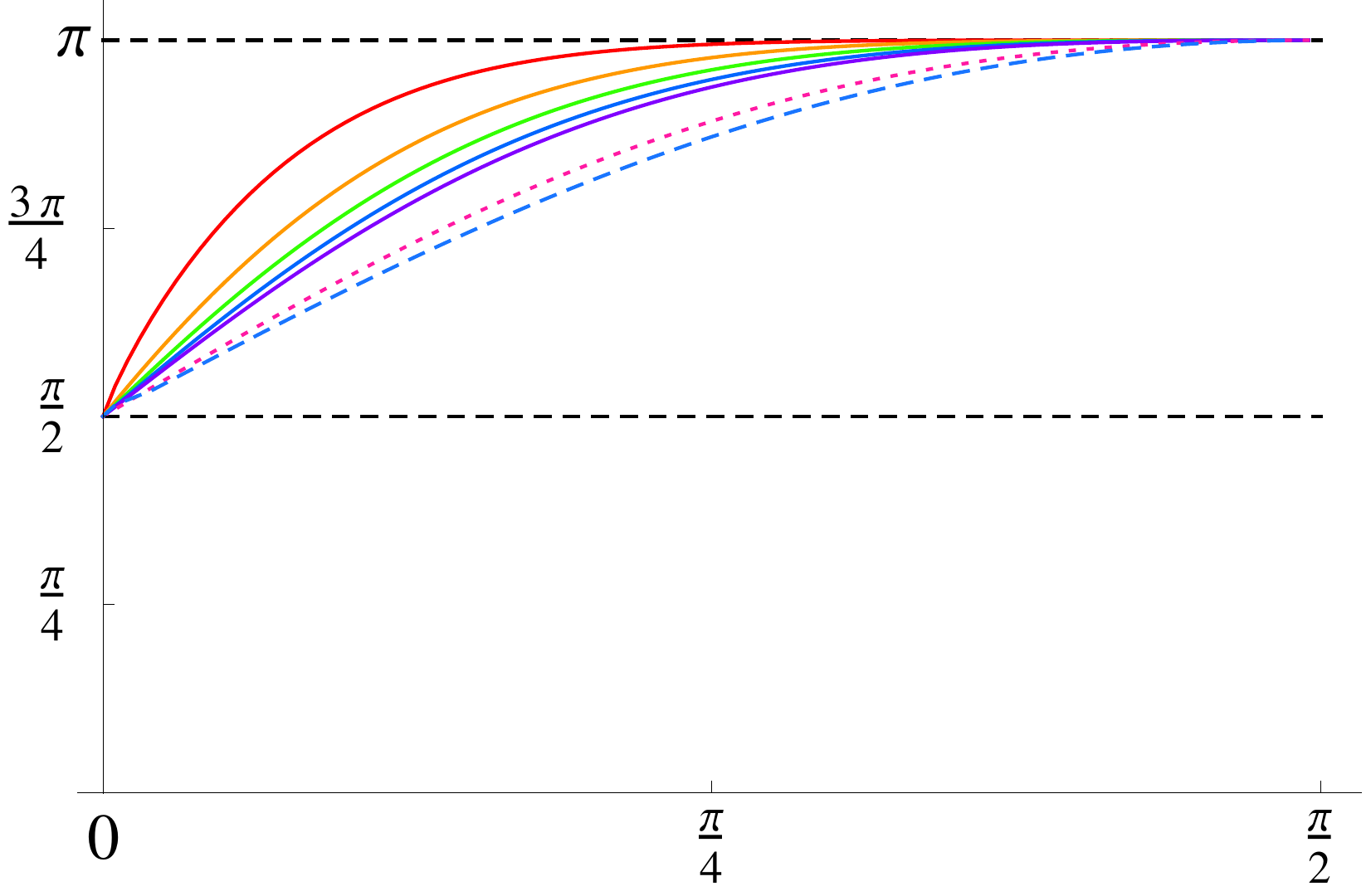} 
\begin{picture}(0,0)
\setlength{\unitlength}{1cm}
\put (-13,8.4) {$\phA^\ast$}
\put (.2,0.7) {$\rho_h$}
\end{picture}
\caption{
The critical curves on $(\phA,\rho_h)$ plane indicating where $\Xi_\CA$ pinches off for \SAdS$_{d+1}$.  $\Xi_\CA$ has two components above the curve and only a single component below.  To guide the eye, we also indicate the $\phA=\pi/2$ and $\phA=\pi$ (dashed lines); the latter gives the upper bound in $\phA$, while the former indicates the lower bound below which $\Xi_\CA$ is connected for \SAdS\ black hole of any size. The topmost (red) curve corresponds to  \SAdS$_4$ geometry where the effect of gravity is strongest, while the next (orange) curve is for \SAdS$_5$ which is our prime exhibit. Increasing the dimension results in slower growth of $\phA^\ast(\rho_h)$, as exemplified by $d=5$ (green), $d=6$ (blue), $d=7$ (purple), $d=19$ (purple dotted) and $d = 49$ (blue dashed). The effects of the weaker gravitational potential are clearly visible with the increasing dimension and the bottommost curve is close to the limiting behaviour for large $d$. }
\label{f:phAvsrhoh}
\end{center}
\end{figure}

We have used this trick to plot the critical curve on $(\phA,\rho_h)$ plane in \fig{f:phAvsrhoh} for various spacetime dimensions. The approximation $\ell_* \approx \ell_0$ can independently be checked by explicit numerical integration and it works extremely well for a large range of black hole sizes.\footnote{
In examples that we have examined we find that $\ell_* - \ell_0 \sim 10^{-3}$ for black holes which are roughly of the order of the AdS radius.}  As expected, for tiny black holes $\rho_h\ll1$, the critical size of the region is $\phA^\ast \to \pi/2$ and grows linearly with $\rho_h \sim \rh$ since there is effectively no other scale in this regime, whereas for very large  black holes $\rho_h\gg1$,  $\phA^\ast \to \pi$ asymptoting to a constant.
Note that as $\phA \to \pi$ we are guaranteed to have a non-trivial topology of $\csf{\cal A}$. In particular, consider the limiting case where ${\cal A} = {\bf S^3}\backslash i^0$, i.e., ${\cal A}$ is a punctured sphere (henceforth denoted as $\phA = \pi^-$). 
Then no matter how large the black hole is, the causal wedge reaches all the way around the boundary while having a hole due to the horizon.
However, from the observation of footnote \ref{fn:horizon} (further discussed in \sec{s:chigen}) it follows that the limit $\phA \to \pi$ is not smooth.

It is  worth remarking that the non-trivial topology of the causal wedge described above relies on working in global \SAdS{} geometry. In the planar \SAdS{} black hole geometry, one cannot circumnavigate the black hole, there being no ``other side''. Nevertheless, as we will describe later, even in the Poincar\'e patch of AdS it is possible to encounter causal wedges with non-trivial topology; for instance  a localized black hole in the Poincar\'e patch will likewise do the trick (see \S\ref{s:ConfSol}).

Above we have presented an example of a causal wedge with one hole.  It is now conceptually easy to generalize this situation to a causal wedge with multiple holes.
For example, we can consider a (dynamical) situation with multiple small black holes in AdS.\footnote{
While such solutions are not known analytically one can construct approximate solutions using a matched asymptotics method.} The black holes will generically orbit each other on timescales set by their separation, radiate gravitational waves, and eventually coalesce.  But we can separate scales in such a way that around each black hole there is a region which is inside a causal wedge for appropriate $\CA$.  In fact, if $\phA \approx \pi/2$, we may see transitions in the number of components of $\Xi$.  For small black holes and $\phA$ larger than $\pi/2$ by amount related to the black hole separation, $\Xi_\CA$ will have a component around each horizon, apart from the one connected to $\partial \CA$.  Hence for a `galaxy' with $N$ separated black holes in AdS, $\Xi$ will have $N+1$ disconnected components.

So far, this section has focused on the topology of the causal wedge and the connectedness of $\Xi_\CA$.  Before closing, let us make an observation about the nature of the `phase transition' between connected and disconnected $\Xi_\CA$ as seen by its area $\chi_\CA$.
Although we don't evaluate the causal holographic information $\chi$ explicitly, we expect that for fixed $\rh$, $\chi_\CA(\phA)$ is not smooth at the transition point $\phAcrit$.  Consider the full curve (on our $(t,\rho,\ph)$ subspace) ${\cal X}_{t=0}$ generated by intersections of future and past geodesics, characterized by $(\rho_{t=0}(\ell),\ph_{t=0}(\ell))$.  The causal information surface $\Xi_\CA$ is a subset of this curve, restricted by $\ph_{t=0}(\ell) \le \pi$; in particular the two curves are identical only when $\Xi_\CA$ has just one connected component.
Since the spacetime is smooth, the geodesics, and hence their intersections, must vary smoothly in $\ell$.  Similarly, the length of the full ${\cal X}_{t=0}$ should vary smoothly as we change $\phA$.  However, at the transition point $\phAcrit$, $\chi$ ceases to be given by length of ${\cal X}_{t=0}$: the difference is given by the finite length piece where $\ph_{t=0}(\ell) > \pi$.  We would therefore expect that $\chi_\CA(\phA)$ has a kink at $\phA = \phAcrit$.

\subsection{BTZ}
\label{s:BTZ}

\begin{figure}
\begin{center}
\includegraphics[width=2in]{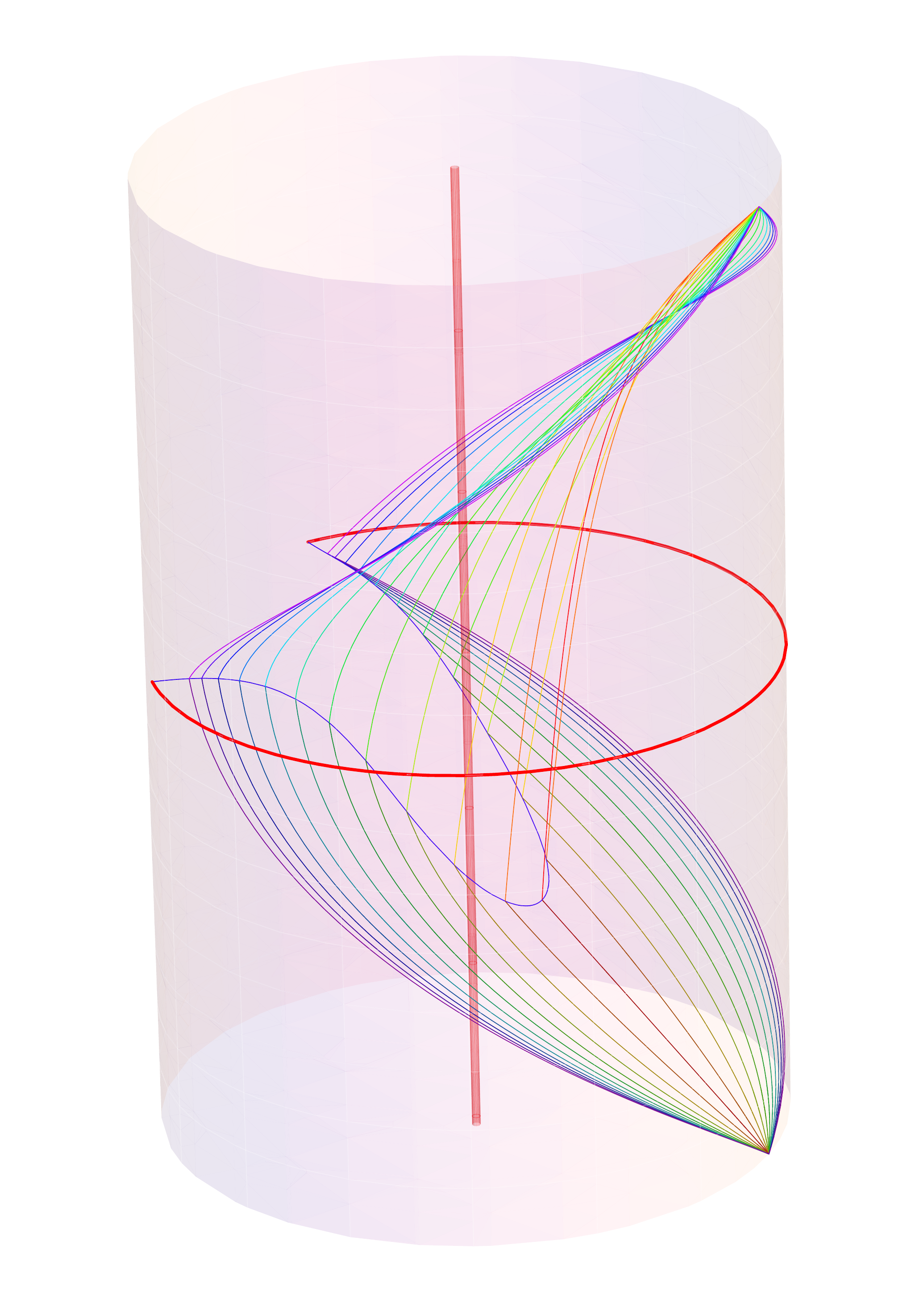} 
\hspace{1cm}
\includegraphics[width=2.5in]{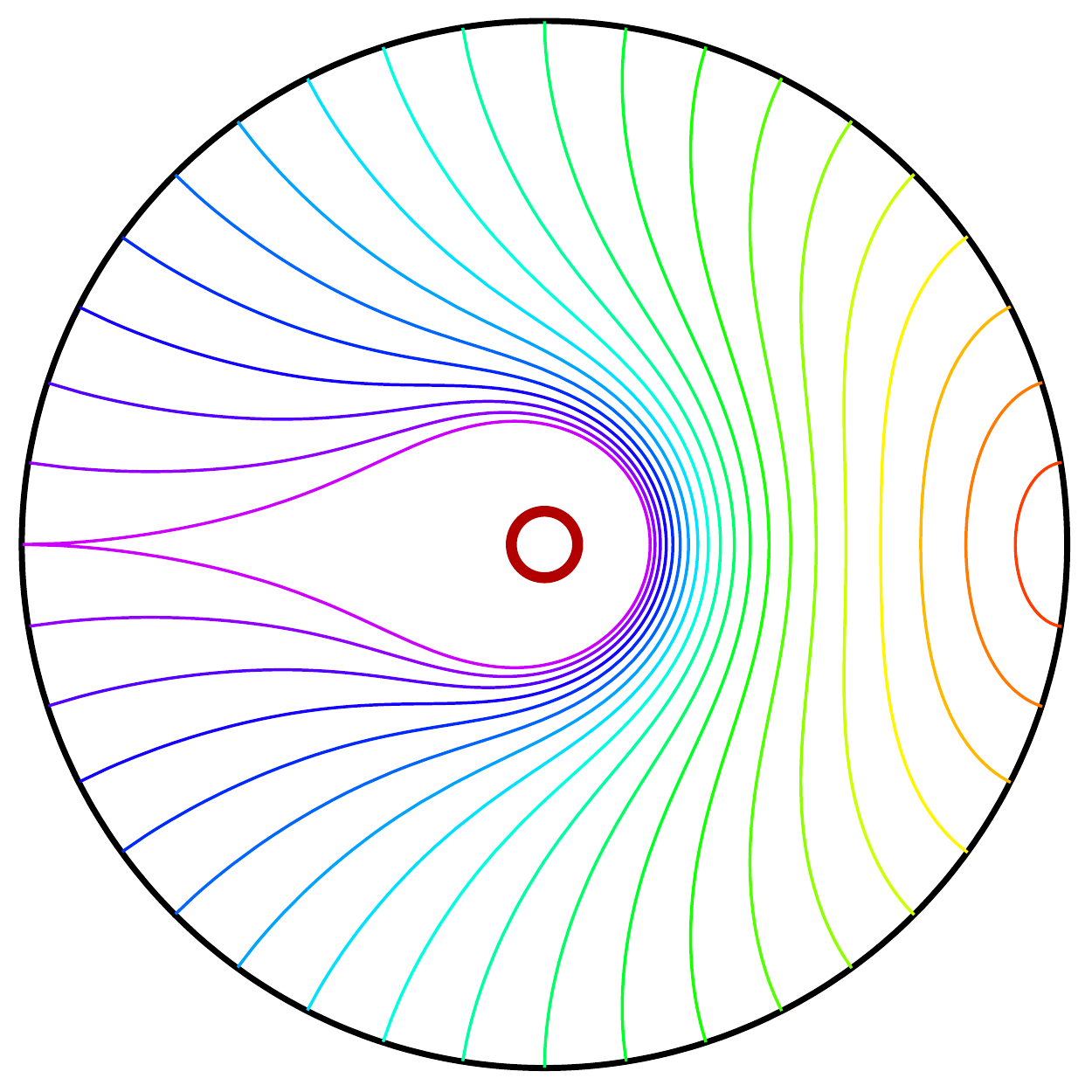}
\caption{
(Left:) A plot of the causal wedge $\cwedge$ and $\Xi_\CA$ (blue curve) in BTZ, with $\rh = 0.02$ and $\phA = 2.5$. 
(Right:)  Projection of spacelike geodesics $\extr{\CA}$ onto the Poincar\'e disk, for varying $\phA \in [0,\pi]$ in increments of $0.05\, \pi$ and $\rh=0.1$.  Since $\extr{\CA} = \Xi_{\cal A}$ in BTZ, the left and right panels are analogous to \fig{f:CWplot} and \fig{f:PDXifoliation}, respectively (modulo the different $\rh$ values).}
\label{f:CWBTZ}
\end{center}
\end{figure}

Having seen the rich structure of $\Xi_\CA$ for \SAdS$_5$, one might wonder whether it is also present in the simpler case of the 3-dimensional BTZ geometry,
which has the form \req{metgofrho} with 
\begin{equation}
g(\rho) = \sin^2 \rho - \rh^2 \, \cos^2 \rho \ .
\label{}
\end{equation}	
Here the calculation is in fact much simpler, and the explicit expressions for  null geodesics which generate $\bCWA$ were presented in e.g.\ \cite{Hubeny:2013hz}.  
A representative causal wedge is plotted  in the left panel of \fig{f:CWBTZ}.  We choose $\phA = 2.5$ as in \fig{f:CWplot}, and we use a tiny black hole $\rh = 0.02$ in order to emphasize the difference from the higher-dimensional case. 
 In fact, as observed already in \cite{Hubeny:2012wa}, in BTZ spacetime $\Xi_\CA$ exactly coincides with $\extr{\CA}$; the latter corresponds to the spacelike geodesic anchored at $\pm \phA$ on the boundary.  The spatial projection of these is described by
\begin{equation}
\tan \rho = \rh \, \frac{\cosh (\rh \, \phA)}{\sqrt{ \cosh^2 (\rh \, \phA) - \cosh^2 (\rh \, \ph) }} \ .
\label{}
\end{equation}	
As can be easily seen, connected spacelike geodesics always exist 
for arbitrary $\phA$, as illustrated in the right panel of \fig{f:CWBTZ}.
 
One reason why the causal wedge does not close off as in the higher dimensional case is that taking the black hole arbitrarily small $\rh \to 0$ does not approach pure AdS: the latter is achieved when $\rh^2 = -1$.  Said differently, in 3 dimensions, the influence of the black hole does not fall off fast enough.  Not only are the effects of a tiny BTZ black hole  perceptible on AdS scale, but even near the boundary there is a qualitative difference between presence and absence of a black hole. A different way to see that 3-dimensional bulk is special is to note that we have argued that causal wedges must be simply connected to the boundary domain ${\cal A}$ \cite{Ribeiro:2007hv} as follows form topological censorship \cite{Galloway:1999bp}. Since in this low dimension $\csf{\cal A}$ is a curve, it must be smoothly deformable to $\domd$. It thus follows that there is no room for non-trivial topology of $\csf{\cal A}$ in three dimensional bulk spacetimes.

\subsection{Boosted black hole}
\label{s:ConfSol}

In \sec{s:SAdS} we have seen that for higher dimensional \SAdS\ black hole of any size, the causal wedge has holes for sufficiently large boundary region $\CA$.  As $\rh \to \infty$, the critical size of $\CA$ for which $\Xi_\CA$ becomes disconnected approaches $\phA^\ast \to \pi$, whereas for $\rh \to 0$, the critical size  $\phA^\ast \to \pi/2$.  In particular, to obtain disconnected $\Xi_\CA$ for this class of geometries, the region $\CA$ must cover at least half of the boundary sphere, and therefore sample a large part of the system.  However, we now argue that large $\CA$ is actually not a prerequisite for existence of disconnected $\Xi_\CA$, in the sense that for {\it any} finite region $\CA$ on the boundary we can construct asymptotically AdS geometries (in more than 3 dimensions) for which $\Xi_\CA$ is disconnected.  

In fact, a simple example which does the job is a boosted version of the global \SAdS\ black hole discussed in \sec{s:SAdS}.  We can consider a family of geometries, considered e.g.\ in \cite{Horowitz:1999gf}, 
corresponding to a boosted global black hole with fixed total energy.\footnote{
See also \cite{Freivogel:2011xc} for a recent discussion where such geometries were called oscillons.}   At zero boost this is the standard global AdS black hole while at infinite boost, this solution limits to a gravitational shock wave in AdS.  In the static coordinates (defined with respect to a specified boundary time), the boosted black hole follows a trajectory which approximates that of a timelike geodesic.

We don't need to do a new calculation to see what will happen for causal wedges in such a geometry, since we can simply boost our causal wedge for the static black hole found in \sec{s:SAdS}.  
In other words, we can implement a coordinate transformation which in pure AdS would transform a timelike geodesic at the origin $\rho(t)=0$ to one which oscillates back and forth with energy $E >1$, whose radial profile is given by
\begin{equation}
\rho(t) = \sin^{-1} \left[ \frac{\sqrt{E^2 -1}}{E} \, \sin t \right] \ .
\label{AdSboostedgeod}
\end{equation}	
The requisite transformation is most easily obtained from isometrically embedding AdS$_{d+1}$ into $\RR^{d,2}$ endowed with the  flat metric $ds^2 = -dX_{-1}^2 - dX_0^2 + \sum_{i=1}^d dX_i^2$, restricted to the hyperboloid  $-X_{-1}^2 - X_0^2 + \sum_{i=1}^d X_i^2 = -1$. This embedding makes the \AdS{} isometries obvious: for instance  we have manifest boost invariance, say $X_0 \to \cosh\beta \, X_0 + \sinh\beta \, X_1$ and $X_1 \to \cosh\beta \, X_1 + \sinh \beta \, X_0$, leaving all the other $X_i$'s unchanged. The energy $E$ in \req{AdSboostedgeod} is related to the boost in the obvious manner $E = \cosh\beta$. So the relevant isometric embedding, which implements the boost and yields AdS in the conformally ESU coordinates, i.e.\ \req{metgofrho} with $g(\rho)=1$, is 
\begin{align}
 X_{0} &= \cosh\beta \, \frac{\sin t}{\cos \rho} + \sinh\beta  \,  \frac{\sin \rho}{\cos \rho} \, \cos \ph \ , &
  X_{-1} &= \frac{\cos t}{\cos \rho} \ ,
\nonumber \\ 
 X_{1} &= \cosh\beta \, \frac{\sin \rho}{\cos \rho} \, \cos \ph + \sinh\beta  \,  \frac{\sin t}{\cos \rho} \ ,&
  X_{k} &=  \frac{\sin \rho}{\cos \rho} \, \sin \ph \; \Omega_k \ , 
\label{AdSisomemb}
\end{align}	
where $\Omega_i$  with $i =2,\cdots, d$ are direction cosines, i.e., $\sum_i \, \Omega_i^2 =1$, coordinatizing a unit ${\bf S}^{d-2}$ and thus explicitly ensuring the $SO(d-1)$ symmetry in the transverse space.
The actual transformation is then generated by comparing the $X$'s for $\beta=0$ with those for arbitrary boost in \req{AdSisomemb}, and is given (modulo some branch issues) by
\begin{align}
& 
{\bar \rho}(\rho,t,\ph) = 
\tan^{-1} \left[ \frac{1}{\cos \rho} \, \sqrt{
(\cosh\beta \, \sin \rho \,  \cos \ph + \sinh\beta \, \sin t)^2
+ \sin^2 \rho \, \sin^2 \ph \,
} \right] 
\nonumber \\ 
& 
{\bar t}(\rho,t,\ph) = 
\tan^{-1} \left[ \cosh\beta \, \tan t + \sinh\beta \, \frac{\sin \rho \, \cos \ph }{\cos t}  \right] 
\nonumber \\ 
& 
{\bar \ph}(\rho,t,\ph) = 
\cot^{-1} \left[ \cosh\beta \, \cot \ph 
+ \sinh\beta \, \frac{\sin t}{\sin \rho\, \sin \ph} \right] 
\label{AdSboostCX}
\end{align}	
where the barred coordinates correspond to boost with respect to the unbarred coordinates by a boost $\beta$. The direction cosines are unchanged since we retain $SO(d-1)$ symmetry.

Note that the boundary is preserved, ${\bar \rho}(\rho=\frac{\pi}{2},t,\ph) = \frac{\pi}{2}$ under \req{AdSboostCX}, and this then specifies the corresponding  transformation induced on the boundary.
More precisely, the bulk coordinate transform implementing the boost corresponds to a conformal transformation (involving both time and space) on the boundary.  This transformation changes the size of $\domd$.  In particular, small $\CA$ and highly boosted black hole translates to large $\CA$ in the static black hole frame, at time where the black hole is closest to the region $\CA$ on the boundary in the boosted picture.  
\begin{figure}
\begin{center}
\includegraphics[width=6.5in]{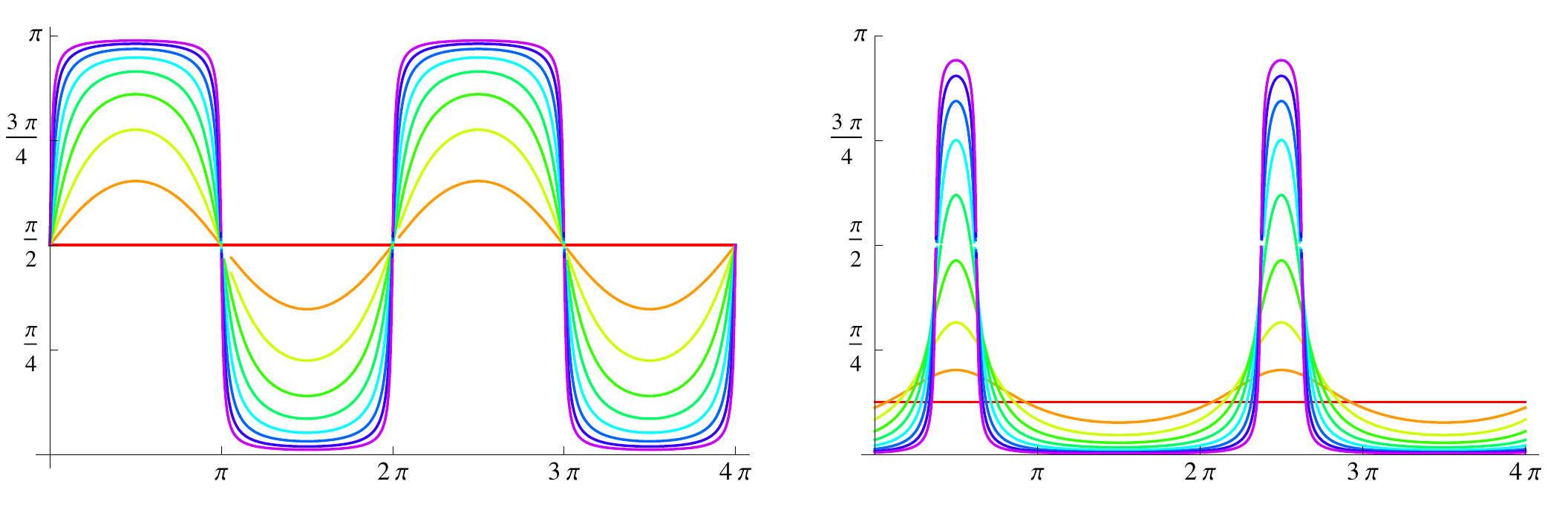} 
\begin{picture}(0,0)
\setlength{\unitlength}{1cm}
\put (-8.3,5.3) {$\phA$}
\put (-16.9,5.3) {$\phA$}
\put (-.3,0.5) {$t$}
\put (-8.6,0.5) {$t$}
\end{picture}
\caption{
Effect of boost on the size $\phA$ of a region $\CA$, as a function of time, as described in the text.  Specifically, we take a fixed-sized region 
${\bar \ph}_\CA = \frac{\pi}{2}$
(Left) and ${\bar \ph}_\CA = \frac{\pi}{8}$ (Right) in the boosted frame and plot how its size $\phA$ varies in the unboosted frame as a function of time $t$, for various values of boost, from unboosted case $\beta = 0$ (red)  to $\beta = 4$ (purple) in increments of 0.5.  This illustrates how arbitrarily small region in boosted frame can look large in unboosted frame for sufficiently large boosts, albeit for time intervals whose size also shrinks with the boost.
}
\label{f:confsolA}
\end{center}
\end{figure}
This is illustrated in \fig{f:confsolA}, which shows how a fixed size region $\CA$ taken at different times translates into a variable-sized region in a boosted frame with relative boost $\beta$ (red curve corresponds to zero boost while purple to $\beta = 4$).  If we take a hemispherical region with ${\bar \ph}_\CA =\pi/2$ (left), the `breathing' of its size $\phA$ in the  boosted frame is symmetric, whereas starting with a small region (right) only produces large sizes in the boosted frame at short intervals.  Nevertheless, it is easy to see that no matter how small we take the region $\CA$ in the unboosted frame, its size in the boosted frame can get arbitrarily close to $\pi$ for sufficiently large boosts.

It is now a simple matter to translate the effect of the boost on the causal wedge: it is precisely the same effect as that of varying the size of $\phA$.  The topological attributes cannot change by the coordinate transformation; so
since  $\cwedge$ has a hole in unboosted frame (equivalently the static black hole spacetime) for large regions, the same is true in the boosted frame (equivalently in the boosted black hole spacetime) even for small regions -- but only at the auspicious times and for sufficient boosts.

It is worth noting that another construction which describes the same (\SAdS) geometry in different coordinates is obtained by simply restricting attention to the Poincar\'e patch of global \SAdS.
The CFT dual on Minkowski space of such configurations was dubbed `the conformal soliton flow' by \cite{Friess:2006kw}.  The event horizon defined with respect to the Poincar\'e patch is then closely analogous to the boundary of the causal wedge, since it is generated by null geodesics in \SAdS\ which end on the boundary at $i^+$ of the Poincar\'e patch which corresponds to some finite ESU time; this was constructed explicitly in \cite{Figueras:2009iu} (cf., their Figure 3 for a plot\footnote{
Strictly speaking the plot in \cite{Figueras:2009iu} is for the BTZ conformal soliton where we have argued for the absence of non-trivial topology; a similar picture in higher dimensions should reveal the appropriate structure.} of the event horizon in global coordinates in 3 dimensions, though the vertical axis is the BTZ time $t$ rather than the coordinate used above in \fig{f:CWBTZ} which is more analogous to Eddington time).  More precisely, the future event horizon indicated there would correspond to $\bcwedgef$ for $\phA = \pi$, though $\Xi_\CA$ for any other $\phA$ can easily be read off by slicing their blue surface horizontally higher than $i^0$.  

So far in this subsection we have discussed geometries which are static though not manifestly so (they all admit a Killing field which is timelike everywhere outside of event horizon). It is even easier to construct examples with disconnected $\Xi_\CA$ for dynamical situations with collapsing and evaporating black holes, where the event horizon lasts only for a finite time.  Again by separation of scales, if the black holes are tiny on AdS scale we can put them anywhere, and they will only eat out tiny spacetime regions from the AdS causal wedge.  Then the original $\bCWA$ (connected to $\partial \domd$) will be only minimally deformed but the full $\bCWA$ will now include additional disconnected components.  This will again translate into $\Xi_\CA$ having multiple components, but now if we try to project them onto a single Poincar\'e disk, they may even intersect.  However, topological censorship will still guarantee that the causal wedge remains simply connected.

Finally, note that since we can take $\CA$ arbitrarily small, the above construction can likewise be implemented in Poincar\'e AdS, where the dual CFT lives on flat space.  In other words, having asymptotically global AdS bulk geometry is not a prerequisite to disconnected $\Xi_\CA$ either.

\subsection{Stars in AdS}
\label{s:genAdS}

So far all our examples of disconnected $\Xi_\CA$ involved causally non-trivial bulk geometries.  In such situations it is easy to argue from first principles that for sufficiently large $\CA$ the causal wedge must have holes, since by causality $\cwedge$ cannot reach past the event horizon.  
However, we will now see that the presence of an event horizon is not a prerequisite for disconnected $\Xi_\CA$.  As long as $\cCA \ne \emptyset$ and we are in more than 3 bulk dimensions, all of the examples discussed above can be modified to keep the geometry causally trivial but nevertheless the causal wedge unaffected.  In black hole geometries, null geodesics starting arbitrarily near the event horizon  take a long time to reach the AdS boundary. So they typically don't make it out to $\domd$; as a result $\Xi_\CA$ remains a finite distance away from the event horizon.\footnote{
This is an extreme case of the time delay effect discussed in \cite{Gao:2000ga}.}
We could then take the metric to remain identical in the region reached by $\cwedge$ but modify it outside that region so as to get rid of all the horizons.  

For example in the \SAdS\ case, we could in principle replace the black hole by a compact star (or even a static shell) which is just slightly bigger than the original black hole.  
To find out whether or not this is physically realistic, let us examine  how compact would such a star  have to be.
Since its maximal size is bounded by the deepest reach of $\Xi_\CA$, and the latter occurs at $\ph=0$, we merely need to see how deep does the radial ($\ell=0$) null geodesic from $\ddtipp$ penetrate by the time $t=0$.  This value of course depends on both $\phA$ and $\rh$.  When $\phA$ is small, the geodesic of course does not have time to reach very deep, but then the causal wedge does not have holes, so this regime is irrelevant for our purposes.  On the other hand, if we take $\phA$ too large in order to guarantee $\cwedge$ having a hole, the geodesics have longer time to travel and typically they approach exponentially close to the horizon.  For example in the case illustrated in \fig{f:CWplot}, the $\rho$ value reached by the radial (red) geodesics $\rho_\Xi$ is already very close to the horizon: $\frac{\rho_\Xi - \rho_h}{\rho_h} \approx 5 \times 10^{-6}$,
which is scarcely realistic for a compact star.

To maximize the size of the requisite star, then, we want to minimize $\phA$ subject to $\Xi_\CA$ being disconnected, i.e.\ take $\phA = \phA^\ast$, and study the corresponding $\rho_\Xi$ as a function of $\rho_h$.
Numerical studies indicate that the maximal value attained, which occurs in the limit of tiny black holes, is $\frac{\rho_\Xi - \rho_h}{\rho_h} \approx 10^{-4}$. 
While this is still not be achievable for physically relevant equations of state for the star, it demonstrates the matter-of-principle point that event horizons are not necessary for topologically non-trivial causal wedge. In fact, preliminary investigations indicate that it might be possible for charged scalar solitons in AdS to have non-trivial causal wedge topology; it would be interesting to explore this further.

\section{General features of $\csf{\cal A}$ and $\chi_{\cal A}$}
\label{s:chigen}

Having seen some curious features of causal wedges in specific classes of examples, we now turn to explaining the more `standard' and completely general properties indicated in \sec{s:preview}.  
We consider generic boundary regions $\CA$ and $\CB$ (using the subscript $t$ to indicate when they lie on the same time slice) and describe simple relational properties\footnote{
See also e.g.\ \cite{Ribeiro:2005wm, Ribeiro:2007hv,Gao:2000ga} for earlier related discussions.
} of causal wedges etc.\ associated with them, including observations pertaining to our constructs in causally non-trivial spacetimes.
We will then compare the areas of various surfaces and consider (sub)additivity properties of $\chi$. 
Finally, we will close with a discussion of extremal surfaces, both within the boundary of the causal wedge and in the full spacetime.
In the process, we will specify a useful relation between $\cwedge$ and any extremal surface  $\extr{\CA}$, whose implications we will consider in \sec{s:extrsurf}. 
 In order to facilitate the reading, in \sec{s:properties} we summarize our claims, leaving their proofs and discussion to \sec{s:proofs}.

\subsection{Summary of properties}
\label{s:properties}

The simple properties which we prove in \sec{s:proofs} are as follows:
\begin{enumerate}[(1).]
\item
If $\CA_t \cap \CB_t  = \emptyset$
(and more generally  if ${\cal{A}}$ and ${\cal{B}}$ are spacelike-separated),
 then $\cwedge$ and $\CW{\CB}$ are spacelike-separated.  
Hence $\cwedge \cap \CW{\CB}  = \emptyset$
and  $\Xi_\CA$ and $\Xi_\CB$  are likewise spacelike-separated.  
\item
If $\CA_t \subset \CB_t$ (and more generally if $\domd \subset \Diamond_{\CB}$), then $\cwedge \subset \CW{\CB}$.  \\
Moreover, if $\CA_t$ is entirely inside $\CB_t$ (more generally if $\partial \domd \cap \partial \Diamond_{\CB} = \emptyset$), then $\Xi_\CA$ and $\Xi_\CB$ are spacelike-separated, with $\Xi_{\CB}$ lying deeper than  $\Xi_{\CA}$ (equivalently outside $\cwedge$).
\item
If $\CA_t$ and $\CB_t$ overlap (i.e.,
if $\CA \cap\CB \ne \emptyset$, but $\CA  \backslash \CB \ne \emptyset$ and  $\CB  \backslash \CA \ne \emptyset$),
 then 
\begin{equation}
\CW{\CA \cap \CB} 
\subset
\CW{\CA} \cap \CW{\CB} 
\subset
\big\{\CW{\CA} , \CW{\CB} \big\}
\subset
\CW{\CA} \cup \CW{\CB} 
\subset
\CW{\CA \cup \CB} \ .
\label{}
\end{equation}	
where by $\big\{\CW{\CA} , \CW{\CB} \big\}$ we mean either $\CW{\CA}$ or $\CW{\CB}$.
Moreover, $\Xi_{\CA \cup \CB}$ and $\Xi_{\CA \cap \CB}$ are spacelike-separated, with $\Xi_{\CA \cup \CB}$ lying deeper than $\Xi_{\CA \cap \CB}$ (i.e.\ outside $\CW{\CA \cap \CB} $), etc.

\item
For any (not necessarily stationary) bulk black hole spacetimes, no causal wedge can penetrate the event horizon.  It then immediately follows that $\Xi_\CA$ cannot penetrate an event horizon for any $\CA$.

\item
Causal holographic information satisfies Additivity: \\
If $\CA \cap \CB = \emptyset$, 
then $\chi_{\CA \cup \CB} = \chi_{\CA} + \chi_{\CB}$.
\item
Causal holographic information satisfies Subadditivity: \\
If $\CA  \cap \CB  \ne \emptyset$, 
then $\chi_{\CA \cup \CB} \le \chi_\CA + \chi_\CB$. \\
(However,  Strong Subadditivity is {\it not} necessarily satisfied, as demonstrated in \cite{Hubeny:2012wa}.)
\item
From the set of all surfaces on $\bCWA$ anchored on $\entsurf$, $\Xi_\CA$ is the minimal-area one.
\item
Extremal surface $\extr{\cal A}$ must lie outside (or on the boundary of) the causal wedge $\cwedge$.\footnote{
However, as demonstrated in \cite{Hubeny:2013hz}, $\extr{\cal A}$ and $\Xi_\CA$ need not be always spacelike-separated.  When they are, the above statement guarantees that $\extr{\cal A}$ lies deeper than $\Xi_\CA$.
\label{fn:vaidyaXiE}}

\end{enumerate}
%

\subsection{Proofs of simple properties}
\label{s:proofs}

Some of the proofs of the above statements use the concept of a causal curve, which we take to be a nowhere-spacelike, maximally extended, connected curve (either in the bulk or along the boundary).
 We will   denote a future-directed causal curve  from a point $p$  to a point $q$ in the spacetime by $\curf {p \to q} = \curp {q \to p}$.   Existence of such a curve guarantees that $q$ is in the future of $p$.

\paragraph{(1). Causal wedges of spacelike-separated regions are spacelike-separated:}

First note that if  ${\cal{A}}$ and ${\cal{B}}$ are spacelike-separated, 
then the corresponding domains of dependence
$\Diamond_{\cal A}$ and $\Diamond_{\cal B}$
 are spacelike-separated.
Otherwise there would exist a (boundary) causal curve which contains points in $\domd$ (and hence, by definition of domain of dependence, must intersect $\CA$) as well as points in $\Diamond_\CB$ (and therefore must intersect $\CB$ as well).  However, existence of causal curve through both $\CA$ and $\CB$  contradicts the assumption that $\CA$ and $\CB$ are spacelike-separated.

We can now extend essentially the same proof-by-contradiction into the bulk: 
If $\cwedge$ and $\CW{\CB}$ are not spacelike-separated, then there exists a (bulk) causal curve $\gamma_{a \to b}$ passing through points $a \in \cwedge$ and $b \in \CW{\CB}$.  w.l.o.g.\ assume that $\gamma_{a \to b} =\gamma^+_{a \to b}$ is future-directed.  By definition of causal wedge, we also know that there exists a causal curve ${\tilde \gamma}_a$ through the point $a$ which starts and ends on the boundary inside $\domd$ and similarly $\exists \ {\Breve {\gamma}}_b$ through $b$ starting and ending in $\Diamond_\CB$.  Out of these three causal curves we can now create a new causal curve 
$\gamma = {\tilde \gamma}^+_{\domd \to a} \cup \gamma^+_{a \to b} \cup {\Breve {\gamma}}^{+}_{b\to \Diamond_\CB}$ 
composed of the past part of   ${\tilde \gamma}_a$, $\gamma^+_{a \to b}$, and future part of ${\Breve {\gamma}}_b$, joined at $a$ and $b$, starting from $\domd$ and ending in $\Diamond_\CB$.   By suitably projecting $\gamma$  onto the boundary obtains a causal\footnote{
The fact that bulk causal curves `project' to boundary causal curves is easy to see: the tangent vector to the bulk causal curve is timelike ($ds^2<0$) and in restricting to the boundary one eliminates a spatial direction which (as long as the projection is performed in such a way as to maintain the relative weighing of temporal and angular components) makes $ds^2$ even more negative.
Also note that we are restricting to the causal wedge of the boundary and so are necessarily outside any black hole horizons.}  curve $\bar \gamma$ intersecting both $\domd$ and $\Diamond_\CB$, contradicting the observation that $\domd$ and $\Diamond_\CB$ are spacelike-separated.
This proves that $\cwedge$ and $\CW{\CB}$ are spacelike-separated.  

In particular, there are no points in $\cwedge$ and $\CW{\CB}$ which can be connected by a causal curve.  Since the causal information surfaces $\Xi$ are contained in the causal wedges, there is correspondingly no causal curve connecting $\Xi_\CA$ and $\Xi_\CB$ -- hence these are likewise spacelike-separated.  Moreover, the absence of causal curve connecting points in  $\cwedge$ and $\CW{\CB}$ trivially implies the absence of common points between  $\cwedge$ and $\CW{\CB}$; in other words,  $\cwedge \cap \CW{\CB} = \emptyset$.   $\square$

\paragraph{(2). Causal wedge inclusion for nested regions:}
First of all, note that if $\CA\subset\CB$, then any boundary causal curve $\bar{\gamma}_a$ through a point  $a \in \domd$ must by definition intersect $\CA$ and therefore it must necessarily also intersect $\CB$, which means that $a \in \Diamond_\CB$.
In other words, $\domd \subset \Diamond_{\CB}$.
We now extend the same argument into the bulk: for any point $a \in \cwedge$, there exists a causal curve $\gamma_a$ which begins and ends in $\domd$ and therefore begins and ends in $\Diamond_{\CB}$.  This implies that $a \in \CW{\CB}$, proving the inclusion $\cwedge \subset \CW{\CB}$. 

To show the rest of the statement, pertaining to $ \domd$ lying strictly inside $ \Diamond_{\CB}$, we first recall an obvious characteristic of points lying on the boundary of a causal wedge, namely the existence of nearby points which lie outside the causal wedge.
In particular, since $\Xi_\CB$ lies on the boundary of $\CW{\CB}$, 
if $p \in \Xi_\CB$, then within any open neighborhood ${\cal O}(p)$, there exists a point $q$ which does not lie within $\CW{\CB}$.
More specifically, since $\Xi_\CB$ lies on the intersection of future and past boundaries of the causal wedge, there exist points $q \in {\cal O}(p)$ through which no causal curve  $\gamma_q$ can start {\it or} end in $\Diamond_\CB$.  This in turn implies that through any $b \in \Xi_{\CB}$, any causal curve $\gamma_b$ can at best make it to $\partial \Diamond_\CB$, but not inside $\Diamond_\CB \backslash \partial \Diamond_\CB$.

We now show that $\Xi_{\CB}$ lies outside $\cwedge$.  
Using the previous observation, we can see that if any point  $b\in \Xi_\CB$ lies inside $\cwedge$, then there exists a causal curve $\gamma_b$ which reaches $\domd \subset \Diamond_\CB \backslash \partial \Diamond_\CB$, a contradiction.
To say this differently, suppose that there exists a point $b \in \Xi_\CB$ which lies inside $\cwedge$.  If $b \in \cwedge \backslash \bCWA$, then there exists an open neighborhood  ${\cal O}(b)$ such that ${\cal O}(b) \subset \cwedge \subset \CW{\CB}$, contradicting the assumption that $b \in \Xi_\CB$.  On the other hand, if no part of $\Xi_\CB$ lies strictly inside $\cwedge$ but $b \in \bCWA$, then $\Xi_\CB$ must be tangent to $\bCWA$.  This means that the generator of $\bCW{\CB}$ through $b$ must coincide with the corresponding generator of $\bCWA$.  Since this generator must extend all the way to the boundary, it terminates on $\partial \Diamond_\CB$ and simultaneously on $\partial \domd$; but this contradicts our assumption that $\partial \domd \cap \partial \Diamond_{\CB} = \emptyset$.  This argument also shows that the boundaries of the two causal wedges cannot coincide at any point, so $\cwedge$ must be strictly inside $\CW{\CB}$.  

Having proved that $\cwedge$ lies strictly inside $ \CW{\CB}$ and that  $\Xi_{\CB}$ lies outside $\cwedge$, the argument that $\Xi_\CA$ and $\Xi_\CB$ must be spacelike-separated proceeds analogously:  w.l.o.g.\, suppose that there exists a future-directed causal curve $\gamma^+_{b \to a}$ (the proof for past-directed curves proceeds analogously) connecting a point $b \in \Xi_\CB$ and a point $a \in \Xi_\CA$.  Then there is a causal curve $\gamma_b = \gamma^+_{b \to a} + {\tilde \gamma}^+_{a \to \domd} $ through $b$ which ends in $\domd \in \Diamond_\CB \backslash \partial \Diamond_\CB$, again a contradiction.
Finally, since $\Xi_\CA$ and $\Xi_\CB$ are spacelike-separated and $\Xi_{\CB}$ lies outside $\cwedge$, it immediately follows that $\Xi_{\CB}$ lies deeper than  $\Xi_{\CA}$.   $\square$

\paragraph{(3). Causal wedge inclusion for overlapping regions:}
Suppose that $\CA$ and $\CB$ overlap,
so that $\CA \cap\CB \ne \emptyset$, with $\CA  \backslash \CB \ne \emptyset$ and  $\CB  \backslash \CA \ne \emptyset$.
Then $\CA \cap \CB$ is a proper subset of 
each of $\CA$ and $\CB$ which are in turn each a proper subset of 
$\CA \cup \CB$, and we can directly apply the results of property (2) discussed above.
In particular, $\Diamond_{\CA \cap\CB} \subset \{ \Diamond_{\CA},\Diamond_{\CB}\} \subset \Diamond_{\CA \cup\CB}$, so that 
$\CW{\CA \cap\CB}  \subset \{ \CW{\CA},\CW{\CB}\}  \subset \CW{\CA \cup\CB}$, and moreover $\Xi_{\CA \cap\CB}$ and $\Xi_{\CA \cup\CB}$  are spacelike-separated, with $\Xi_{\CA \cup \CB}$ lying deeper than $\Xi_{\CA \cap \CB}$.

To see the `intermediate' inclusions, it is evident that 
$\CW{\CA} \cap\CW{\CB}  \subset \{ \CW{\CA},\CW{\CB}\}  \subset \CW{\CA} \cup \CW{\CB}$, and all that remains to show is that 
$\CW{\CA \cap\CB}  \subset \CW{\CA} \cap\CW{\CB}$ and 
that $\CW{\CA} \cup \CW{\CB} \subset \CW{\CA \cup\CB}$.
The arguments are similar to the ones used above: if $p \in \CW{\CA \cap\CB} $, then there exists a causal curve $\gamma_p$ which starts and ends in $\Diamond_{\CA \cap\CB} $, so it necessarily also starts and ends in each of $\Diamond_{\CA} $ and $\Diamond_{\CB} $, and therefore $p$ lies in both $\cwedge$ and $\CW{\CB}$ -- which implies that $p \in \CW{\CA} \cap\CW{\CB}$.  The other inclusion is similarly manifest.   $\square$

\paragraph{(4). Causal wedge cannot penetrate event horizon:}
This statement follows immediately from the definition of an event horizon:  since there is no future-directed causal curve from inside the black hole which can make out it to the AdS boundary, no point inside the black hole can lie inside the causal wedge of any boundary region.  Correspondingly, the causal information surface $\Xi_\CA$ for any sub-region $\CA$ of the total boundary space must lie outside the event horizon.\footnote{
As an aside, note that this property of remaining outside the black hole does not generically hold for extremal surfaces, due to the teleological nature of the event horizon: an extremal surface cannot be sensitive to its exact location  in dynamically evolving spacetimes, and therefore can probe inside the black hole; explicit examples have already been seen in \cite{AbajoArrastia:2010yt,Hubeny:2013hz} and will be further discussed in \cite{Hubeny:2013uq}.  (In contrast, as pointed out in \cite{Hubeny:2012ry}, in static spacetimes extremal surfaces don't penetrate the event horizon either.
Nevertheless, any extremal surface $\extr{\CA}$ must lie outside a causal wedge $\cwedge$ as we show in Property 8.)
}

In asymptotically global AdS spacetimes, there is  a slight subtlety:  as we take $\CA$ to cover the entire spatial section of the boundary, $\domd$ jumps discontinuously from having finite time-extent (given by the size of the boundary sphere) to having infinite time extent (and covering the entire boundary spacetime).  In this special case the causal wedge $\cwedge$ is an open set,  its boundary coincides with the event horizon, and $\Xi_\CA$ then lies along the event horizon bifurcation surface.  
On the other hand, for $\CA$ being a proper subset of the boundary Cauchy slice, the finiteness of $\domd$ implies that $\Xi_\CA$ can reach only to within a finite (albeit quantitatively small) distance from the horizon; we saw an example in \sec{s:genAdS}.

\paragraph{}
Now that we have discussed the relational properties between the causal wedges and the causal information surfaces for two regions, let us briefly turn to the causal holographic information $\chi$.  Recall that this quantity is potentially the one most directly accessible from the field theory.

\paragraph{(5). Additivity:}
If $\CA \cap \CB = \emptyset$, then by property (1), 
$\cwedge \cap \CW{\CB}  = \emptyset$, so $\Xi_\CA$ and $\Xi_\CB$ are disjoint.  Hence  $\Xi_{\CA \cup \CB}= \Xi_\CA \cup \Xi_\CB$, and the individual areas then simply add up,
 $\chi_{\CA \cup \CB} = \chi_{\CA} + \chi_{\CB}$.
 Note that each of the terms is divergent, with LHS inheriting all the divergences from RHS.

\paragraph{(6). Subadditivity:}
If $\CA  \cap \CB  \ne \emptyset$, then $\CA  \cup \CB$ is not simply composed of two disjoint regions as in property (5) above.  Instead, the surface area of this region is strictly smaller than the sum of the individual surface areas, ${\rm Area}(\partial[\CA  \cup \CB]) < {\rm Area}(\partial[\CA]) + {\rm Area}(\partial[\CB])$, because the RHS  has additional contribution from ${\rm Area}(\partial[\CA  \cap \CB])$.  Now, since $\chi_\CA$ has its leading divergence proportional to ${\rm Area}(\partial[\CA])$ and similarly for other regions, 
the Subadditivity inequality $\chi_{\CA \cup \CB} \le \chi_\CA + \chi_\CB$ is satisfied trivially -- i.e.\ the RHS has stronger divergence.
On the other hand, as noted in \sec{s:properties}, the property of Strong Subadditivity, where the leading divergences cancel, is actually not necessarily satisfied.

\paragraph{}
Properties (5) and (6), along with the basic feature of being (quarter of) the proper area of a co-dimension two surface, make it tempting to compare the causal holographic information $\chi_\CA$ with the entanglement entropy $S_\CA$.  
The failure of $\chi_\CA$ to satisfy strong subadditivity is the strongest evidence that it cannot correspond to a von Neumann entropy.

Finally, let us close with two properties which have been observed previously, and deal with extremal surfaces.  

\paragraph{(7). $\Xi_{\cal A}$  is a minimal surface on $\bCWA$:}
We can in fact view this property as an alternate definition of $\Xi_\CA$, which is conceptually useful for considering the causal holographic information $\chi_\CA$.  However, to establish the result, we use the definition \req{defXi}, that $\Xi_\CA$ belongs to both $\bcwedgef$ and $\bcwedgep$.
The proof assumes null energy condition and uses a crucial observation about null congruences which generate the causal wedge boundary: the null congruences $\partial_{\pm}(\cwedge)$  must have non-negative (and generically positive) expansion in the outgoing (towards the boundary) direction.  For if they had negative expansion, Raychaudhuri equation would imply that they caustic before reaching the AdS boundary, in contradiction to them generating the boundary of a causal past/future of $\domd \in \bdy$.  

Let us first restrict attention to the case where $\cwedge$ is topologically trivial, i.e.\ $\Xi_\CA$ is a single connected surface anchored on $\partial \CA$.
Consider any other surface $\Upsilon_{\cal A} \subseteq \bCWA$ which is anchored on $\partial \CA$.  We want to show that $\Upsilon_{\cal A}$ cannot have smaller area than $\Xi_\CA$.  We can obtain a subset\footnote{
Recall that new generators can enter $\bCWA$ at caustics, so if $\Upsilon_{\cal A}$ lies closer to the boundary than the first caustic, all generators from $\Xi_\CA$ pass through $\Upsilon_{\cal A}$ but in addition some new ones do as well.
} of  $\Upsilon_{\cal A}$ from  $\Xi_{\cal A}$ by flowing a certain distance $\lambda$ along the null generators.  Let us for the moment assume that  $\Upsilon_{\cal A} \in \bcwedgef$; then we can perform a constant rescaling of the affine parameter of each null generator individually, such that $\Upsilon_{\cal A}$ lies at constant affine parameter $\lambda_0$ along the null generators of $\bcwedgef$.
Now, using the fact that the expansion of the null generators of $\bcwedgef$ cannot be negative towards the boundary, we know that the area of constant $\lambda$ slices of $\bcwedgef$ must be monotonically increasing function of $\lambda$; in particular,
${\rm Area}(\Upsilon_{\cal A}) \ge {\rm Area}(\Xi_{\cal A}) $.  
(The presence of caustics in $\bcwedgef$ would only strengthen this inequality, since the area of the remaining part of $\Upsilon_{\cal A}$, which does not lie along generators from $\Xi_\CA$, is positive.)
Same argument would apply for $\Upsilon_{\cal A}$ lying on $\bcwedgep$, as the past-directed null generators again expand towards the boundary.
If $\Upsilon_{\cal A}$ lies partly on $\bcwedgef$ and partly on $\bcwedgep$,  then we can separate $\csfz{\cal A}$ into domains, separated by $\Upsilon_{\cal A} \cap \Xi_{\cal A}$, and run the argument for each domain separately.  Hence in all cases, any  surface $\Upsilon_{\cal A}$ cannot have smaller area than $ \csfz{\cal A}$, which means that $\Xi_{\cal A}$ is the minimal surface on $\bCWA$. 

Let us now turn to the topologically non-trivial case, such as discussed in \sec{s:SAdS}, where $\cwedge$ has a hole and $\Xi_\CA$ consists of several disconnected surfaces.  Then the statement of minimality of $\Xi_\CA$ has to be made more precise, as clearly there are surfaces $\Upsilon_{\cal A} \subseteq \bCWA$ with $\partial \Upsilon_{\cal A} = \partial \CA$ with smaller area than $\Xi_\CA$ -- as a trivial example, take just the connected part of $\Xi_\CA$ anchored on $\partial \CA$, or a small deformation thereof.  However, if  in addition to $\partial \Upsilon_{\cal A} = \partial \CA$ we require that $\Xi_{\CA}$ and $\Upsilon_{\cal A}$ are homotopic to each other within $\bCWA$,
then our arguments above go through:  any generator of $\bCWA$ which starts from $\Xi_\CA$ must intersect $\Upsilon_{\cal A}$, where the positive expansion of the generators guarantees that the latter has larger area. $\square$

The fact that $\Xi_{\cal A}$  is a minimal (and therefore extremal)  surface on $\bCWA$ however does not mean that it is an extremal surface in the full bulk spacetime, as emphasized previously in \cite{Hubeny:2012wa,Hubeny:2013hz}.  The final property will show this explicitly, by providing a relational property between $\extr{\CA}$ and $\cwedge$.

\paragraph{(8). Extremal surface is generically outside causal wedge:}
This is a generalization of the proof presented in \cite{Hubeny:2012wa} (see also \cite{Wall:2012uf}), which 
was formulated on a given spacelike slice of the bulk rather than the full Lorentzian geometry.\footnote{
In that more limited context, one can equivalently say that the extremal surface $\extr{\CA}$ reaches deeper than the causal information surface $\Xi_\CA$; however, in general highly dynamical spacetimes such statement need not be meaningful when  $\extr{\CA}$ and $\Xi_\CA$ are not spacelike-separated.
}
  The reasoning here is similar; the crux is to argue that if the extremal surface came within the causal wedge, it would have to be tangent to the boundary of a nested causal wedge corresponding to some sub-region.  Once this is established, we can obtain a contradiction from comparing the expansions for the extremal surface and a tangent slice of a causal wedge boundary.
  
Suppose that an extremal surface $\extr{\CA}$ has a sub-region $e$ which lies strictly inside $\cwedge$.
We claim that then somewhere along $e$, there exists a point $p$ at which $\extr{\CA}$ is tangent to $\bCW{\CB}$ corresponding to some smaller boundary region ${\cal B}\subseteq \CA$.
There may in fact be infinitely many of such points, but our proof only requires one.  The existence of one follows easily if one can foliate $\cwedge$ by $\bCW{\alpha}$ for a  family of nested regions $\CA_\alpha$, parameterized by  some parameter $\alpha$, which we can choose to be $\alpha=0$ at $\bCWA$ and monotonically growing as the size of $\CA_\alpha$ decreases.\footnote{
For example, one convenient parameterization is by the depth to which $\Xi_\alpha$ reaches, $\alpha = r_{\Xi_\alpha} - r_{\Xi_\CA}$.  Alternately, for spherically symmetric spacetimes such as considered in \sec{s:SAdS}, we can simply take $\alpha = \phA - \ph_\alpha$.
}
For then this foliation specifies a function $\alpha$ on $\extr{\CA}$ (given by the particular $\bCW{\alpha}$ intersected by our surface $\extr{\CA}$ at any specified point).  Since $\alpha = 0$ at $\extr{\CA} \cap \bCWA$, i.e.\ on $\partial e$, and positive inside $e$, there is a point $p\in e$ where $\alpha$ takes extremal value.  At this point $p$, $\extr{\CA}$ is tangent to $\bCW{\alpha}$.\footnote{
A potential loophole is that  $\bCW{\alpha}$ might be non-smooth at the point of intersection with $\extr{\cal A}$, e.g,  $p$ could be one of the cusps  seen in \fig{f:PDXifoliation}. In such cases there are two potential ways to extend our arguments. One is that we could consider other foliations of ${\cal A}$ on the boundary to ensure that the intersection of the extremal surface with the causal wedge boundary is at a regular point. More simply however, we could also note that even at the cusps the expansion of the null congruence towards the boundary while divergent is manifestly positive definite (and we only need the correct sign for our argument to go through). We thank Matt Headrick and Aron Wall for useful discussions on this point.  }

Since the above argument followed from the existence of a foliation of $\cwedge$ by $\bCW{\alpha}$, we now pause to briefly consider how generally does such a foliation exist.
First of all, it is not hard to see that for static spacetimes, a foliation of $\cwedge$ is indeed guaranteed to exist; as a canonical example (which includes caustics and potentially non-trivial topology of $\cwedge$), consider the \SAdS{} example discussed in \sec{s:SAdS}.  Since the projections of $\Xi_\CA (\phA)$ on the Poincar\'e disk for different $\phA$ indicated in  \fig{f:PDXifoliation} foliate the entire region of the Poincar\'e disk which is in the largest causal wedge, this spatial foliation lifts trivially to the full Lorentzian region inside the largest causal wedge.\footnote{
Here we envision that $\domd$ is compact on the boundary. The issue is more subtle if we take ${\cal A}$ to be a complete Cauchy slice of the boundary  ESU; as noted in footnote \ref{fn:horizon}, $\domd$ is the entire boundary spacetime and its causal wedge does not admit such a foliation (as apparent from both \fig{f:PDXifoliation}  and \fig{f:CWBTZ} (right panel)). }  
More generally, we can argue that such a foliation must exist for any static spacetime as follows.
Let us w.l.o.g.\ fix $\CA$ to lie at $t=0$ on the boundary.
We first pick a convenient foliation of the region $\CA$ by
 slicing $\partial \domd$ by constant $t = \alpha$ surfaces and projecting to the $t=0$ slice.  Denote each leaf of $\CA$'s foliation by $\alpha$ and the enclosed subregions by $\CA_\alpha$.  In case of circular regions, $\alpha \in (0,\phA)$. 
Notice that the causal wedge $\CW{\alpha}$ for a given sub-region $\CA_\alpha$ can be  obtained by simply rigidly sliding $\bcwedgef$ and $\bcwedgep$ towards each other in time.
Now consider some point $p \in \cwedge$.  Since there is a unique point in $\partial_\pm(\cwedge)$ whose temporal projection coincides with that of $p$, $\exists ! \, \alpha$ by which we can slide $\partial_\pm(\cwedge)$ such that $p \in \partial \CW{\alpha}$.
In other words, this rigid time-translation of $\partial_\pm(\cwedge)$  towards each other defines a natural foliation of $\cwedge$.
We believe that it is possible to extend this proof to more general spacetimes, although we leave a rigorous argument for future investigation.

Let us now return to the main argument, having established that if a region $e \subset \extr{\CA}$ lies within $\cwedge$, then there exists a point $p \in e$ at which $\extr{\CA}$ is tangent to $\bCW{\CB}$ for some sub-region $\CB$.
In fact, we can construct a spacelike surface $\Psi_\CB$ within $\bCW{\CB}$ which is tangent to $\extr{\CA}$ at $p$ and is anchored on $\partial \CB$.
At this point, outgoing null normal to $\extr{\CA}$ coincides with the outgoing null normal to $\Psi_\CB$ (which is  the corresponding generator of $\bCW{\CB}$).  
Now consider the expansions $\Theta_{\extr{}}$ and $\Theta_\Psi$ of the two tangent surfaces $\extr{\CA}$ and  $\Psi_{\CB}$ at $p$.  By definition of extremal surface, we know $\Theta_{\extr{}} = 0$.  On the other hand, since $\partial \CB$ is a boundary of a causal set, $\Theta_\Psi \ge 0$ towards the boundary.  This is however a contradiction, since the way in which these surfaces are tangent to each other (with $\extr{}$ bending away from the boundary more than $\Psi$) implies that $\Theta_{\extr{}} > \Theta_\Psi$.  The reader is encouraged to consult Fig.5 of \cite{Hubeny:2012wa} for a pictorial sketch of this argument.

Since the assumption that $\extr{\CA}$ reaches inside $\cwedge$ produced a contradiction, we conclude that $\extr{\CA}$ must lie  outside (or at best on the boundary of) $\cwedge$.  $\square$

In the generic situation where $\Xi_\CA$ is spacelike-separated from $\extr{\CA}$, it then immediately follows that $\extr{\CA}$ reaches deeper than $\Xi_\CA$.
However, having established that an extremal surface $\extr{\CA}$ cannot lie within the causal wedge, we should note that this does not automatically imply that $\extr{\CA}$ is necessarily causally disconnected from $\domd$ (although this is the case in generic situations, when $\extr{\CA}$ and $\Xi_\CA$ are spacelike separated).  Apart from the obvious special examples where  $\extr{\CA}$ and $\Xi_\CA$ coincide, as pointed out in footnote$^{\ref{fn:vaidyaXiE}}$, in the case of thin shell Vaidya-AdS explored in \cite{Hubeny:2013hz} we saw that they can be null-separated.

\section{Implications for bulk extremal surfaces}
\label{s:extrsurf}

So far, we have been discussing the causal wedge and related constructs in the bulk spacetime, which as yet have no independently defined construction in the dual CFT.  In this section we point out that our results nevertheless bear on a more familiar context where we {\it do} have a conjectured duality.  In particular, they surprisingly turn out to be relevant for the entanglement entropy $S_\CA$, which is conjectured  \cite{Ryu:2006bv,Ryu:2006ef,Hubeny:2007xt} to be given by (quarter of) the area of the extremal surface $\extr{\CA}$ which (i) is anchored on the entangling surface, $\partial \extr{\CA} = \partial \CA$,  (ii) is homologous to $\CA$, and (iii) in case of multiple such surfaces is the minimal-area one.

In the previous section, we have further justified the anticipated result that no extremal surface $\extr{\CA}$ can lie within the causal wedge $\cwedge$ (Property 8). 
While we have presented this as a property of causal wedges, we can conversely think of it as a property of extremal surfaces.  As already mentioned in \sec{s:intro} the consequence that extremal surfaces penetrate deeper into the bulk than causal wedges bears on the question of how much of the bulk does a given boundary region in the CFT describe.  But quite apart from this discussion, there is a more remarkable and surprising consequence of Property 8, when combined with the observation of \sec{s:CWtopol} that causal wedges can have holes:  in a global eternal black hole spacetime, if the causal wedge $\cwedge$ for a given  boundary region $\CA$ has a hole, then there {\it cannot} exist a connected extremal surface $\extr{\CA}$ anchored on $\partial \CA$ which is homologous to $\CA$.  

To see why this is the case, let us first consider the homology requirement.   Suppose we take a region $\CA$ which covers more than half of the boundary, such as the case indicated in \fig{f:CWplot}.  If area minimization was the only constraint on the desired extremal surface $\extr{\CA}$ anchored on $\partial \CA$, then the extremal surface passing around the opposite side of the black hole from $\CA$ (i.e.\ through $\ph=\pi$) would be the relevant one; let us denote it as $\extr{\CA^c}$.  However, such a surface is not homologous to $\CA$, since there does not exist a co-dimension one bulk smooth hypersurface whose only boundary are $\extr{\CA}$ and $\CA$ -- such hypersurface would have to pass through the black hole, and  would either encounter the black hole curvature singularity, or pass through the Einstein-Rosen bridge, in which case it would have further boundaries.  The upshot is that, to satisfy the homology requirement, we should either take a surface $\extr{\CA}$ which goes on $\CA$'s side of the black hole (i.e.\ passes through $\ph=0$), or a pair of disconnected extremal surfaces, $\extr{\CA^c}$ and the bifurcation surface $\CH$ of the event horizon $r=\rh$.

Now let us consider what happens when the causal wedge has a hole, as in  \fig{f:CWplot}.  
A connected spacelike\footnote{
Since the bulk geometry is static, in this case of an eternal black hole we can in fact w.l.o.g.\ take the extremal surface (which is anchored at constant $t$ on the boundary) to lie at constant $t$ in the bulk.  (In more complicated geometries such as discussed in \cite{Freivogel:2005qh} where there could be more general extremal surfaces, the homology requirement would no longer rule out $\extr{\CA^c}$  \cite{Hubeny:2013ee}, so the present consideration would be irrelevant.)
} 
surface anchored at $\partial \CA$ would then either have to stay at larger radial position than the connected part of $\Xi_\CA$ (which is ruled out by the homology requirement), or it must pass around the black hole, in which case it must enter the causal wedge.  To justify the latter more formally, in order to pass through $\ph=0$ and remain outside the causal wedge there, $\extr{\CA}$ would have to attain smaller radial value than the deepest reach of the disconnected part of $\Xi_\CA$, so it would have to pass through the radial region traversed by the causal wedge caustics (${\cal C}^\pm$ in \fig{f:CWplot}).  Since this radial region is contained within the causal wedge at all angles, this suffices to guarantee the passage though the causal wedge.  But since that would violate Property 8,  no such connected extremal surface $\extr{\CA}$ homologous to $\CA$ can exist. $\square$

Comparing this result to the behaviour of spacelike geodesics in BTZ (cf.\ the right panel of \fig{f:CWBTZ}), the non-existence of requisite extremal surfaces in higher dimensions might seem rather surprising -- in fact, we are not aware of this effect having been noticed previously.  Given that these surfaces exist only for sufficiently small $\phA$, and reach deeper into the bulk as $\phA$ increases,
one might naturally wonder what happens as we try to push  this deepest reach $r_{\extr{}}$ further towards the horizon.  This is examined in detail in \cite{Hubeny:2013ee}, with the curious result that the corresponding $\phA$ does not behave monotonically; instead it oscillates between certain maximal and minimal values.  Even more curiously, $\phA(r_{\extr{}})$ oscillates infinitely many times as  $r_{\extr{}} \to \rh$, exhibiting a self-similar behaviour.  In other words, for a fixed region $\CA$ with $\phA$ in a certain range, there are in fact infinitely many extremal surfaces (all outside the causal wedge) anchored on $\partial \CA$ and homologous to $\CA$.  But conversely for large enough $\phA$, there are none.

Let us now ask what are the implications of these results for the entanglement entropy.  As discussed in  \cite{Hubeny:2013ee}, we learn that the entanglement entropy cannot be a smooth function of $\phA$.  At some critical $\phA$, the relevant surface which determines entanglement entropy switches from the $\extr{\CA}$ to the $\extr{\CA^c}+\CH$ family, at which point $S(\phA)$ has a kink and saturates onto a plateau.
This phenomenon has been described as holographic entanglement plateau in \cite{Hubeny:2013ee}, whose Fig.9 shows  the various critical curves for $\phA(\rh)$ in the case of \SAdS$_5$ black holes. We note that these authors choose to display the result in terms of $\alpha = \frac{1}{\pi}\left(\phA - \sin\phA \,\cos\phA\right)$; for the casual wedges this is a rescaled version of our \fig{f:phAvsrhoh}. We must emphasize that the critical point $\phA(\rh)$ for the entanglement plateau transition of \cite{Hubeny:2013ee} is only weakly determined by the casual wedge topology (indeed the presence of such an effect has been widely argued for since the inception of the holographic entanglement entropy proposal \cite{Ryu:2006bv, Headrick:2007km}). This has to do with the fact that while non-trivial topology in the causal wedge is sufficient for the entanglement plateaux to develop, it certainly is not necessary. To ascertain the onset of the plateau phenomenon one requires detailed dynamics of minimal surfaces, and indeed as in any first order phase transition the exchange of dominance between the $\extr{\CA}$ to the $\extr{\CA^c}+\CH$ families occurs before it is mandated by causal wedge topology.

\section{Discussion}
\label{s:discuss}

We have examined global properties of causal wedges and related constructs, proving a number of useful relational statements valid in arbitrary causal asymptotically AdS bulk spacetime.  While from the CFT standpoint, the causal wedge might have been hitherto viewed as a rather esoteric construct, the preceding section demonstrated that its properties bear on more familiar CFT quantities such as the entanglement entropy.  Nevertheless, the ultimate goal of this exercise is to use these properties to try to propose or actually construct the CFT dual of the causal wedge or associated quantities.  The relational statements we discuss in \sec{s:chigen} are physically quite reasonable, and indeed unsurprising; the most curious feature, analyzed in \sec{s:CWtopol}, is the topologically non-trivial nature of the causal wedge $\cwedge$ for simple regions $\CA$.
 
Bolstered by the remarkable ease with which we have been able to produce `holes in the causal wedge', the reader might well wonder whether  this is perhaps the generic situation, namely whether in {\it any} bulk geometry satisfying the genericity condition, we could identify {\it some} region $\CA$ for which $\Xi_\CA$ would have disconnected components.   This would have strong implications for any putative CFT dual.

Since for the \SAdS$_{d+1}$ family of solutions, the case which was guaranteed to have disconnected $\Xi_\CA$  was the one with maximal $\CA = {\bf S}^{d-1}\backslash i^0$, i.e.\ $\phA = \pi^-$, let us examine this situation for   general bulk geometries.  
What is the generic form of $\Xi_\CA$ and $\cwedge$? 
Note that domain of dependence\footnote{
Typically domain of dependence $\Diamond_{\Sigma}$ is defined for a closed achronal surface $\Sigma$; whereas in the present case $\CA$ is open.  However, since $\Diamond_{\Sigma}$ is defined as the set of all points from which every {\it timelike} curve intersects $\Sigma$, it should still be true in the present case that $\Diamond_{\Sigma}$ is closed.
} $\domd$ in this case 
coincides with the AdS Poincar\'e wedge boundary, and in particular  includes the Poincar\'e wedge spatial infinity vertex $i^0$.  Moreover the tips $q^{\wedge,\vee}$ of $\domd$ correspond to $i^\pm$ of the Poincar\'e wedge.  This in turn implies that $\cwedge$ does include $i^0$, and since this point lies both on $\bcwedgep$ and $\bcwedgef$ (joined to $q^{\wedge,\vee}$ by the boundary null geodesics), it is contained in $\Xi_\CA$.  Now the question is, is $\Xi_\CA$ just that one point $i^0$, or does it reach into the bulk?  In other words, are there points on bulk Cauchy slice anchored at $\tA$ which don't lie in $\cwedge$?

For black holes the answer is clearly yes; but even for causally trivial spacetimes, one might naively expect that the answer is generically {\it yes}, with pure AdS being the only exception.  The motivation for this expectation is as follows: consider a near-boundary point $p$ just radially in from $i^0$.  In AdS, $J^\pm(p)$ intersects the boundary within $\domd$ only near $q^{\wedge,\vee}$.  When we excite the CFT state by a bulk deformation (satisfying the null energy condition), these bulk geodesics are time-delayed \cite{Woolgar:1994ar,Gao:2000ga} so that they don't make it to $\domd$.
Hence $p$ would seem not to lie in $\cwedge$; but since $\cwedge$ does contain the entire boundary sphere, it should have a hole around $p$.  This would lead us to expect that $\Xi_\CA$ generically reaches into the bulk (presumably in some teardrop-like shape anchored on the single boundary point $i^0$).

However, there is a subtlety with this reasoning, having to do with the presence of caustics in $\bCWA$. It is easy to see that radial null geodesics are time-delayed by some finite amount when traversing the bulk with some gravitational potential well.  However, these are irrelevant, because generically they exit $\bCWA$ and enter inside the causal wedge at a caustic point formed by null geodesics with `angular momentum'\footnote{
Angular momentum along geodesics is a well-defined constant of motion only in presence of orbital Killing field; but as a warm-up to address part of the genericity question, we will focus on spherically symmetric spacetimes in this discussion.
} $\pm \varepsilon$, for some $\varepsilon \ll 1$. The geodesics which we need to focus on are the ones with maximal angular momentum.
 These are repelled by the centrifugal potential and as such stay close to the boundary feeling thereby less influenced by the deformation in the core (IR) region of the geometry.\footnote{
We saw this behaviour in \fig{f:CWplot}, but 
one can also verify this  explicitly in more general static spherically symmetric examples: the larger angular momentum geodesics are less time delayed than the smaller angular momentum ones (including  the radial one), so they will remain on $\bCWA$ longer, covering up the smaller-angular momentum geodesics under a seam of caustics.}   This suggests that the relevant geodesics to consider are the near-boundary geodesics with angular momentum $\ell = \pm (1-\varepsilon)$.  

Null geodesics through deformed AdS (for a static spherically symmetric metric corresponding to a ``star" geometry) were examined by \cite{Hubeny:2006yu} in the context of bulk cone singularities.  Let us consider a null congruence emanating from a specified point $\ddtipp$ on the boundary.  In pure AdS, each null geodesic in this congruence would reach the boundary at the antipodal point, with\footnote{
Note that  \cite{Hubeny:2006yu} considered equatorial geodesics with polar angle $\ph\in[0,2\pi)$, whereas here we are using $\ph\in[0,\pi]$ to represent the azimuthal angle.  However, as explained in \sec{s:SAdS} due to spherical symmetry the geodesic equations are of course identical, so we can conflate the two for direct comparison.
} $\Delta \ph = \pi$ and $\Delta t  = \pi$.  When the geometry has a gravitational potential well, the geodesics exhibit the usual time-delay and light-bending effect, which leads to a deformation of the future endpoint $(\Delta \ph,\Delta t)$  of each geodesic, parameterized by the reduced angular momentum $\ell \in (0,1)$.  
A parametric plot of $\Delta t (\Delta \ph)$ of geodesic endpoints for typical static spherically symmetric deformations of AdS were plotted in \cite{Hubeny:2006yu} (cf., their Fig. 5); the slope of the curve (parameterized by $\ell$) in the $(\Delta \ph,\Delta t)$ plane representing null geodesic endpoints is simply $\ell$.  
This means that in general, near-boundary null geodesics reach the boundary at a more spatially than temporally shifted endpoint, $\Delta \ph > \Delta t$.  This would seem to imply that the $\pm \ell$ geodesics in the same (future or past) congruence intersect each other `before' (i.e.\ closer to $\domd$) than they intersect the corresponding $\ell$ geodesic from the other congruence.  

One can check these expectations explicitly in a given deformed bulk geometry.  For example, consider the bulk metric \req{metgofrho} with $g(\rho) = 1 - \nu \, \cos^4 \rho$ with $\nu \ll 1$.  This is a causally trivial asymptotically AdS 
geometry with a gravitational potential well given by $\nu$, with pure AdS corresponding to $\nu = 0$.  The causal wedge indeed remains topologically trivial, with a seam of caustics reaching all the way to $i^0$.  In fact, this is to be expected from the black hole result of \sec{s:SAdS}: the shape of (the connected component of) $\Xi_\CA$ for large enough $\phA$ should be determined only by the asymptotic geometry, and should therefore be the same for corresponding to either a black hole or a star of the same mass.  From \fig{f:PDXifoliation} we see that as $\phA \to \pi$ (cf.\ the purple curves on the left), $\Xi_\CA$ does {\it not} approach a tear-drop shape.  In particular, without the disconnected component around the horizon, $\Xi_\CA$ would indeed simply retract to $i^0$ as $\phA \to \pi$.  So we expect that in order to obtain causal wedges with non-trivial topology, the bulk spacetime has to be sufficiently deformed from pure AdS.

Said another way, there is an interesting tension between two competing effects. Had the time-delay effect been the only operative feature, we would have concluded that holes in the causal wedge would have been generic. On the other hand, if centrifugal repulsion which forces attention on near-boundary geodesics were to be the primary effect, there would be no room for non-trivial topology. It is then clear that the natural place to look for non-trivial causal wedge topology is when the two effects are competing, and indeed our prototypical example of \SAdS{} provides one such setting.  Naively then, we can abstract from the above discussion an essential requirement for the causal wedge to develop holes: {\em the spacetime must admit null circular orbits}.\footnote{
Of course, this can only be a requirement for spherically symmetric spacetimes, where such orbits may exist in the first place.  In absence of any symmetries, we would have to formulate a more robust criterion.  
} For it is in this case that there is some non-trivial interplay between the two effects discussed above and they precisely offset each other at the circular orbit. Indeed we saw that the critical transition point for \SAdS{}
was effectively determined by the null circular orbit in \S\ref{s:SAdS}. 
The justification provided there leads us to conjecture that in spherically symmetric spacetimes the presence of the null circular orbits is a necessary condition for non-trivial causal wedge topology.
It would be useful to prove this statement rigorously;  we hope to return to this interesting problem in the near future. 
 
One lesson of our explorations which we wish to emphasize is that  while studies of global AdS contain the Poincar\'e AdS case (as a limit), the converse is not true.  Not only is the global case much richer (as can be expected already from the metric being more complicated), but it gives us novel insight into important observables. Over the years we have learnt many interesting lessons by examining field theories on compact spatial geometries and we believe that there is indeed much more to be learned from similar explorations. 

\acknowledgments 
It is a pleasure to thank Matt Headrick and Harvey Reall useful discussions. MR and VH would like to thank CERN, ITF Amsterdam, ICTP and Crete Center for Theoretical Physics for hospitality during the course of this project.  MR would like to thank the organizers of 
the ``Relativisitc fluid dynamics and the gauge gravity duality'' workshop held at Technion, Haifa and those of the ``2nd Mediterranean Conference on Classical And Quantum Gravity'', Veli Losinj, Croatia for  their hospitality during the concluding stages of this project.  ET would like to thank the NBI Copenhagen and Physics dept.\ of the University of Turin for hospitality during part of this work.
VH and MR are supported in part by the the STFC Consolidated Grant ST/J000426/1.



\providecommand{\href}[2]{#2}\begingroup\raggedright\endgroup

\end{document}